\documentclass[11pt,a4paper]{article}
\usepackage{jheppub}
\pdfoutput=1

\usepackage{color}
\usepackage[utf8]{inputenc}
\usepackage{amsmath}
\usepackage{amsfonts}
\usepackage{tikz}
\usetikzlibrary{arrows}
\usetikzlibrary{shapes}
\usepackage[english]{babel}
\usepackage[autostyle]{csquotes}
\usepackage{braket} 
\usepackage{xcolor}
\usepackage{tabularx}

\usepackage{graphics}
\usepackage{tkz-euclide}
\usepackage[toc,page]{appendix}
\usetikzlibrary{decorations.markings,arrows}
\usetikzlibrary{arrows,shapes,backgrounds}
\usetikzlibrary{decorations.pathreplacing,decorations.markings}
\usetikzlibrary{calc, trees, positioning, arrows, shapes, shapes.multipart, shadows, matrix, decorations.pathreplacing, decorations.pathmorphing}
\usetikzlibrary{shapes.misc}
\usepackage{hyperref}
\usepackage{bookmark}
\usepackage{verbatim}
\usetikzlibrary{shapes.geometric}
\usepackage{tikz-cd}
\usepackage{thmtools, thm-restate}

\newcommand{\CC}{\mathcal C}

\newcommand{\CN}{\mathcal N}

\newcommand{\qu}{\mathbb H}

\newcommand{\be}{\begin{equation}} \newcommand{\ee}{\end{equation}}
\newcommand{\eref}[1]{(\ref{#1})}
\newcommand{\bigslant}[2]{{\raisebox{.2em}{$#1$}\left/\raisebox{-.2em}{$#2$}\right.}}

\def\c3d#1{\mathcal{C}^{3d}(#1)}

\def\node#1#2{\overset{#1}{\underset{#2}{\circ}}}

\def\ba{\begin{equation} \begin{aligned}} \def\ea{\end{aligned}\end{equation}}

\preprint{Imperial/TP/18/AH/07}
\title{Discrete Gauging in Coulomb branches of Three Dimensional $\CN=4$ Supersymmetric Gauge Theories}
\author[a]{Amihay Hanany}
\author[a]{, Anton Zajac}
\affiliation[a]{Theoretical Physics, The Blackett Laboratory\\
Imperial College London\\ SW7 2AZ United Kingdom}
\emailAdd{a.hanany@imperial.ac.uk}
\emailAdd{anton.zajac@imperial.ac.uk}

 \abstract{This paper tests a conjecture on discrete non-Abelian gauging of $3d$ $\CN=4$ supersymmetric quiver gauge theories.
Given a parent quiver with a bouquet of $n$ nodes of rank $1$, invariant under a discrete $S_n$ global symmetry, one can construct a daughter quiver where the bouquet is substituted by a single adjoint $n$ node. Based on the main conjecture in this paper, the daughter quiver corresponds to a theory where the $S_n$ discrete global symmetry is gauged and the new Coulomb branch is a non-Abelian orbifold of the parent Coulomb branch. We demonstrate and test the conjecture for three simply laced families of bouquet quivers and a non-simply laced bouquet quiver with $C_2$ factor in the global symmetry.}


\begin{document}
\maketitle

\newpage
\section{Introduction} \label{0}

A $3d$ $\mathcal{N}=4$ quiver gauge theory typically has a Coulomb branch of the moduli space that is a
hyperk\"ahler singularity \cite{CHZ14,N15,B15,BR16}. The graph theoretical nature of such quivers opens a large field for the study of orbifolding and other actions of discrete groups \cite{C17}. This paper is devoted to a particular action on quivers which has both, a gauge theoretic, as well as a geometric interpretation.

We consider a large class of quivers which have a set of $U(1)$ nodes attached to a common pivot node. Apart from this \emph{complete bouquet}\footnote{\emph{Complete} means that all the nodes of the bouquet are of rank $1$. Generically, bouquet can consist of nodes of any ranks, not necessarily the same within the bouquet.} of nodes, the rest of the quiver is arbitrary as the statement to be made on the quiver is a purely local one. The action on the quiver can be summarized by taking the set of $n$ $U(1)$ nodes and replacing them with an adjoint $n$ node. The construction is formulated by Conjecture \ref{Main Conjecture}, which is a more general version of Conjecture 1 in \cite{HZ18}. 

\newtheorem{thm}{Conjecture}
\begin{thm}[Discrete Gauging]
\label{Main Conjecture}

Given a 3d $\CN=4$ quiver ${\sf Q}_{\{1^n\}}$\footnote{The partition notation for bouquet quivers is explained later in this section.} with $n$ nodes of rank $1$ attached to another node of rank $k$, (gauge node or global node\footnote{The special case when the pivot node is a global flavor node is discussed in section 4.1 in \cite{CHZ14}.}) (Fig.\eref{fig:1}),
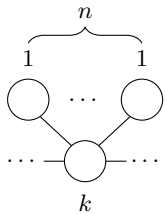
\begin{figure}[h!]
\label{bouquet}
\center{
\begin{tikzpicture}[scale=0.68]
\draw (0.4,0) -- (0.8,0);
\draw (0,0) circle (0.4cm);
\draw (0,-0.8) node {\footnotesize{$k$}};
\draw (1.2,0) node {\footnotesize{$\dots$}};
\draw (-0.4,0) -- (-0.8,0);
\draw (-1.2,0) node {\footnotesize{$\dots$}};
\draw (-0.26,0.3) -- (-0.87,0.87);
\draw (-1.1,1.2) circle (0.4cm);
\draw (-1.1,2) node {\footnotesize{$1$}};
\draw (0.26,0.3) -- (0.87,0.87);
\draw (1.1,1.2) circle (0.4cm);
\draw (1.1,2) node {\footnotesize{$1$}};
\draw (0,1.2) node {\footnotesize{$\dots$}};
\draw [decorate,decoration={brace,amplitude=6pt}] (-1.1,2.3) to (1.1,2.3);
\draw (0,2.9) node {\footnotesize{$n$}};
\end{tikzpicture}
}
\caption{\label{fig:1} ${\sf Q}_{\{1^n\}}$ quiver.}
\end{figure}
one can construct a new 3d $\CN=4$ quiver ${\sf Q}_{\{n\}}$ with an adjoint $n$ node attached to $k$ (Fig.\eref{fig:2}).
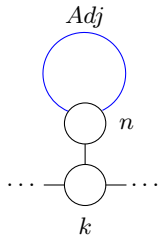
\begin{figure}[h!]
\center{
\begin{tikzpicture}[scale=0.68]
\draw (0.4,0) -- (0.8,0);
\draw (0,0) circle (0.4cm);
\draw (0,-0.8) node {\footnotesize{$k$}};
\draw (1.2,0) node {\footnotesize{$\dots$}};
\draw (-0.4,0) -- (-0.8,0);
\draw (-1.2,0) node {\footnotesize{$\dots$}};
\draw (0,0.4) -- (0,0.8);
\draw (0,1.2) circle (0.4cm);
\draw (0.8,1.2) node {\footnotesize{$n$}};
 \draw[-,blue] (0.335,1.455) arc (-64:247:0.8);
 \draw (0,3.27) node {\footnotesize{$Adj$}};
\end{tikzpicture}
}
\caption{\label{fig:2} ${\sf Q}_{\{n\}}$ quiver.}
\end{figure}
Then, the following relation \eref{eq:00} between the Coulomb branches of these quivers holds
\begin{equation} \label{eq:00}
\mathcal{C} \left ( {\sf Q}_{\{n\}} \right ) = \mathcal{C} \left({\sf Q}_{\{1^n\}}  \right ) / S_n ~,
\end{equation}
where $S_n$ is the discrete symmetry group of permutations of $n$ elements.
\end{thm}
The quiver in Figure \eref{bouquet} has a natural $S_n$ symmetry which permutes the $U(1)$ gauge nodes, and the corresponding Coulomb branch inherits this symmetry as a discrete global symmetry. A natural step in a geometric construction of moduli spaces is to gauge a subgroup of the discrete global symmetry, resulting in a new moduli space. From gauge theoretical perspective, one constructs a new theory, given by the quiver in Figure \eref{fig:2}, such that the Coulomb branches satisfy Equation \eref{eq:00} of Conjecture \eref{Main Conjecture}. 

Discrete gauging has been ascribed physical interpretation for a particular class of $6$d $\mathcal{N}=(1,0)$ supersymmetric theories that describe low energy physics of a set of $n$ M5 branes on a $\mathbb{C}^2/ \mathbb{Z}_k$ singularity in \cite{HZ18}. The Higgs branch of such theories at infinite coupling can be expressed as a Coulomb branch of a $3d$ $\mathcal{N}=4$ quiver gauge theory. System of $n$ separated M5 branes on $\mathbb{C}^2/ \mathbb{Z}_k$ singularity, has a discrete $S_n$ global symmetry on the moduli space. This arises from the manifest permutation symmetry of the corresponding $M5$ branes (i.e. the positions of the separated M5 branes). By making some of the M5 branes coincident a subgroup of the discrete global symmetry $H_{\lambda} \subseteq S_n$ is gauged. $H_{\lambda}$ corresponds to a partition $\lambda$ that describes subsets of M5 branes that are coincident. For every partition, different $H_{\lambda}$ is gauged, producing a theory with a Coulomb branch that is a non-Abelian orbifold of the parent Coulomb branch ($\mathcal{C}_{\lambda},\; \lambda=1^n$), which corresponds to $n$ separated M5 branes on a $\mathbb{C}^2/ \mathbb{Z}_k$ singularity.

In this paper, Conjecture \eref{Main Conjecture} applies to $3d$ $\CN=4$ quivers that need not necessarily describe low energy dynamics of systems of M5 branes on ALE singularities. In the present work, Conjecture \eref{Main Conjecture} describes a phenomenon purely in $3$d without any reference to other dimensions.\\
 
In order to test and provide evidence of Conjecture \eref{Main Conjecture}, we study a class of quivers which contain a sub-quiver such as depicted in Figure \eref{bouquet}, consisting of a \emph{bouquet} of $n$ rank $1$ gauge nodes that stems from a rank $k$ gauge node. Appendix \ref{A} contains an example of a simple construction that can be employed to obtain bouquet quivers starting from a generic unitary quiver with flavors. In this paper, we perform the discrete gauging construction prescribed in Conjecture \eref{Main Conjecture} for quivers with unitary gauge nodes, however, it should be emphasized that analogous construction can be formally defined for much broader class of quivers \footnote{Strictly speaking, one only requires a presence of a bouquet without any additional requirements on the node from which the bouquet stems. In particular, the pivot node can be an ortho-symplectic (i.e. $O$, $SO$ or $Sp$) gauge node.}.\\

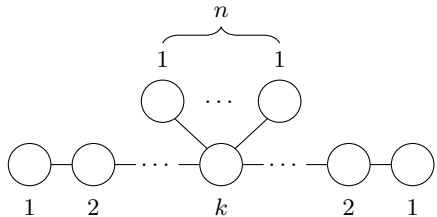
\begin{figure}[h!]
\center{
\begin{tikzpicture}[scale=0.70]
\draw (0.4,0) -- (0.8,0);
\draw (0,0) circle (0.4cm);
\draw (0,-0.8) node {\footnotesize{$k$}};
\draw (1.2,0) node {\footnotesize{$\dots$}};
\draw (2.4,0) circle (0.4cm);
\draw (3.6,-0.8) node {\footnotesize{$1$}};
\draw (2.4,-0.8) node {\footnotesize{$2$}};
\draw (-2.8,0) -- (-3.2,0);
\draw(3.6,0) circle (0.4cm);
\draw (-2.4,0) circle (0.4cm);
\draw (-3.6,-0.8) node {\footnotesize{$1$}};
\draw (-2.4,-0.8) node {\footnotesize{$2$}};
\draw (-0.4,0) -- (-0.8,0);
\draw (1.6,0) -- (2,0);
\draw (2.8,0) -- (3.2,0);
\draw (-3.6,0) circle (0.4cm);
\draw (-1.2,0) node {\footnotesize{$\dots$}};
\draw (-1.6,0) -- (-2,0);
\draw (-0.26,0.3) -- (-0.87,0.87);
\draw (-1.1,1.2) circle (0.4cm);
\draw (-1.1,2) node {\footnotesize{$1$}};
\draw (0.26,0.3) -- (0.87,0.87);
\draw (1.1,1.2) circle (0.4cm);
\draw (1.1,2) node {\footnotesize{$1$}};
\draw (0,1.2) node {\footnotesize{$\dots$}};
\draw [decorate,decoration={brace,amplitude=6pt}] (-1.1,2.3) to (1.1,2.3);
\draw (0,2.9) node {\footnotesize{$n$}};
\end{tikzpicture}
}
\caption{\label{fig:4} $A$-type quiver with $\mathcal{P}_{[1^n]}(n)$ bouquet.}
\end{figure}

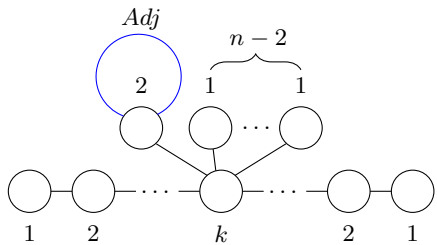
\begin{figure}[h!]
\center{
\begin{tikzpicture}[scale=0.70]
\draw (0.4,0) -- (0.8,0);
\draw (0,0) circle (0.4cm);
\draw (0,-0.8) node {\footnotesize{$k$}};
\draw (1.2,0) node {\footnotesize{$\dots$}};
\draw (2.4,0) circle (0.4cm);
\draw (3.6,-0.8) node {\footnotesize{$1$}};
\draw (2.4,-0.8) node {\footnotesize{$2$}};
\draw (-2.8,0) -- (-3.2,0);
\draw(3.6,0) circle (0.4cm);
\draw (-2.4,0) circle (0.4cm);
\draw (-3.6,-0.8) node {\footnotesize{$1$}};
\draw (-2.4,-0.8) node {\footnotesize{$2$}};
\draw (-0.4,0) -- (-0.8,0);
\draw (1.6,0) -- (2,0);
\draw (2.8,0) -- (3.2,0);
\draw (-3.6,0) circle (0.4cm);
\draw (-1.2,0) node {\footnotesize{$\dots$}};
\draw (-1.6,0) -- (-2,0);
\draw (-0.26,0.3) -- (-1.22,0.9);
\draw (-1.5,1.2) circle (0.4cm);
\draw (-1.5,2) node {\footnotesize{$2$}};
\draw (0.26,0.3) -- (1.22,0.9);
\draw (1.5,1.2) circle (0.4cm);
\draw (1.5,2) node {\footnotesize{$1$}};
\draw (-0.2,1.2) circle (0.4cm);
\draw (-0.2,2) node {\footnotesize{$1$}};
\draw (-0.1,0.39) -- (-0.15,0.8);
\draw (0.7,1.2) node {\footnotesize{$\dots$}};
\draw [decorate,decoration={brace,amplitude=6pt}] (-0.2,2.3) to (1.5,2.3);
\draw (0.7,2.9) node {\footnotesize{$n-2$}};
 \draw[-,blue] (-1.2,1.455) arc (-64:249:0.8);
 \draw (-1.5,3.27) node {\footnotesize{$Adj$}};

\end{tikzpicture}
}
\caption{\label{fig:5} $A$-type quiver with $\mathcal{P}_{[2,1^{n-2}]}$ bouquet.}
\end{figure}

To establish the notation, consider the bouquet quiver in Figure \eref{fig:4}, and let us make the following remarks. Firstly, in a computation of the Coulomb branch, a center of mass $U(1)$ always decouples. All the Coulomb branches in this paper are computed by decoupling the $U(1)$ on the central node since for simply laced quivers, the Coulomb branch does not change if one decouples the $U(1)$ on a different gauge node of the quiver\footnote{A non-simply laced quiver has, in general, $K$ different Coulomb branches, where $K=\#_s +1$. $\#_s$ is the number of short nodes modulo outer automorphisms of the quiver.}. Secondly, various arrangements and ranks of the bouquet nodes of a given quiver are in one-to-one correspondence with partitions of $n$. In particular, in Figure \eref{fig:4} and Figure \eref{fig:5} the upper nodes are arranged in the form of $\mathcal{P}_{[1^n]}(n)$ and $\mathcal{P}_{[2,1^{n-2}]}(n)$, respectively. Throughout this paper we use partition notation to describe the arrangements and ranks of the bouquet nodes. \\

In order to perform discrete gauging, let us gauge $H$, a subgroup of the discrete symmetry $S_n$, that acts on the bouquet of nodes in Figure \eref{fig:4}. For $H=\mathbb{Z}_2$, according to Conjecture \eref{Main Conjecture}, one obtains the daughter quiver depicted in Figure \eref{fig:5}.

The bouquet of the new quiver in Figure \eref{fig:5} consists of $n-2$ copies of $U(1)$ nodes and a single $U(2)$ node with an adjoint loop. The adjoint loop adds extra hypermultiplet contributions to the conformal dimension $\Delta$ (\cite{BK02,GW09}, (2.4) in \cite{CHZ14}) of BPS operators\footnote{For detailed discussion of BPS operators, see f.i. \cite{BBHY07}.} that live in that particular node. The addition of extra hypermultiplets is straightforwardly adjusted for, and implemented, in the \emph{monopole formula}, (2.7) in \cite{CHZ14}, used for the computation of the Coulomb branch. Examples of quivers with adjoint nodes recently appeared in \cite{MOTZ17,HMinstanton2018}.\\

It is not yet espied which families of quivers with bouquets are the most interesting for their physical or mathematical properties. From the possible landscape of bouquet quivers the following three families are studied in the present paper:

\begin{itemize}
  \item Star-shaped quivers with a central $2$ node and a bouquet of $1$ nodes
  \item Quivers consisting of a chain of $n_2$ rank $2$ nodes with two bouquets:\\
  The first bouquet with $n_1$ rank $1$ nodes is attached to the leftmost chain node\\
   The second bouquet with two rank $1$ nodes is attached to the rightmost chain node
    \item $A$-type quivers with outer $\mathbb{Z}_2$ automorphism symmetry and a bouquet that stems from the central node\footnote{Compare with (2.12) in \cite{HZ18}.}
\end{itemize}

Figure \eref{fig:4} and \eref{fig:5} show examples of quivers which belong to the third family. We can parametrize this family by $n$ and $k$. For $k=2$ one recovers the first family of quivers. Considering the quivers belonging to the second family and setting $n_2 = 1$ (i.e.  if the \enquote{chain} contains only a single $2$ node), one also recovers the first family.

\begin{figure}[h!]
\center{
\begin{tikzpicture}[scale=0.70]
\draw (0.4,0) -- (0.8,0);
\draw (0,0) circle (0.4cm);
\draw (0,-0.8) node {\footnotesize{$k$}};
\draw (1.2,0) node {\footnotesize{$\dots$}};
\draw (2.4,0) circle (0.4cm);
\draw (3.6,-0.8) node {\footnotesize{$1$}};
\draw (2.4,-0.8) node {\footnotesize{$2$}};
\draw (-2.8,0) -- (-3.2,0);
\draw(3.6,0) circle (0.4cm);
\draw (-2.4,0) circle (0.4cm);
\draw (-3.6,-0.8) node {\footnotesize{$1$}};
\draw (-2.4,-0.8) node {\footnotesize{$2$}};
\draw (-0.4,0) -- (-0.8,0);
\draw (1.6,0) -- (2,0);
\draw (2.8,0) -- (3.2,0);
\draw (-3.6,0) circle (0.4cm);
\draw (-1.2,0) node {\footnotesize{$\dots$}};
\draw (-1.6,0) -- (-2,0);
\draw (0,0.4) -- (0,0.8);
\draw (0,1.2) circle (0.4cm);
\draw (0.8,1.2) node {\footnotesize{$n$}};
 \draw[-,blue] (0.335,1.455) arc (-64:247:0.8);
 \draw (0,3.27) node {\footnotesize{$Adj$}};
\end{tikzpicture}
}
\caption{\label{fig:6} $A$-type quiver with $\mathcal{P}_{[n]}(n)$ bouquet.}
\end{figure}
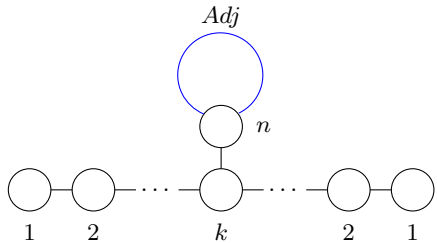

Starting with Figure \eref{fig:4}, one can draw quivers for all partitions of $n$. For each $U(r)$ node with $r>1$ in the bouquet, one remembers to add an adjoint loop. By gauging the entire global $S_n$ symmetry of the theory in Figure \eref{fig:4}, and using Conjecture \eref{Main Conjecture}, one obtains the quiver in Figure \eref{fig:6}, corresponding to the last partition $\mathcal{P}_{[n]}(n)$.
It is natural to study the relations between Coulomb branches corresponding to the various partitions $\mathcal{P}(n)$. 
Let $\mathcal{P}_{[\lambda]}(n)$ be a daughter theory constructed from the parent complete bouquet quiver $\mathcal{P}_{[1^n]}(n)$. Conjecture \eref{Main Conjecture} implies that the Coulomb branches satisfy
\be \label{eq:XX}
\mathcal{C}_{[\lambda]} = \mathcal{C}_{[1^n]}/\Gamma,
\ee
where $\Gamma$ is a discrete symmetry group that corresponds to the difference of the global permutation symmetry between the complete $\mathcal{P}_{[1^n]}(n)$ bouquet and the $\mathcal{P}_{[\lambda]}(n)$ bouquet, respectively. Equation \eref{eq:XX} has the following implication on the volumes of the two Coulomb branches. One can expand the unrefined Hilbert series around the $t=1$ pole
\be \label{eq:expansionHS}
HS(t) \mid_{t\rightarrow 1} \sim \frac{R}{(1-t)^d},
\ee
where $d$ is the complex dimension of the Coulomb branch and $R$ denotes the value of the residue at the pole. Then, since Equations \eref{eq:vol1} and \eref{eq:vol2}
\begin{gather}
vol(\mathcal{C_{[\lambda]}})= R_{\lambda} \label{eq:vol1} \\
vol(\mathcal{C}_{[1^n]})= R_{1^n} \label{eq:vol2}
\end{gather}
define the volumes of the Coulomb branches, Equation \eref{eq:01} is satisfied.
\be \label{eq:01}
\frac{vol(\mathcal{C}_{[1^n]})}{vol(\mathcal{C}_{[\lambda]})} = \frac{R_{1^n} }{R_{\lambda}} = ord(\Gamma)
\ee
Note, that $ord(\Gamma)$ denotes the order of the discrete group $\Gamma$. In this note, the discrete gauging construction of Conjecture \eref{Main Conjecture} is applied to all three aforementioned families of quivers. As a result, for all possible gauged subgroups $H_{\lambda} \subseteq S_n$ of the discrete global symmetry, one can study the obtained Coulomb branches and perform a collection of non-trivial tests verifying that the daughter Coulomb branches are non-Abelian orbifolds of the parent Coulomb branch. The same construction is done for a particular representative of non-simply laced quivers with $C_2$ factor in the global symmetry. One of the motivations for including non-simply laced quivers is merely to emphasize that non-simply laced theories are equally important to study as the simply laced gauge theories. The comparison of the Coulomb branch volumes of the unrefined Hilbert Series is used as a necessary non-trivial test of Equation \eref{eq:00}. Direct comparison of the refined Hilbert Series can be used for an exact verification of Conjecture \eref{Main Conjecture}. For the latter, one needs to study how the refined Hilbert series of theory $\mathcal{P}_{[1^n]}(n)$ maps to that of the $\mathcal{P}_{[\lambda]}(n)$ theory. In particular, one can use the character maps between the corresponding character expansions of the Hilbert series. \\

\subsection{The layout of the paper}

In section \ref{1} we set the stage by performing an analysis for the \emph{first family} of bouquet quiver theories. The analysis has the following structure.

\paragraph{The Analysis}

Assuming that Conjecture \eref{Main Conjecture} holds, for each partition (corresponding to different gauging of the discrete global symmetry), we display the corresponding quiver alongside with its imbalance and quaternionic dimension. The imbalance of the unbalanced node as well as the quaternionic dimension of the Coulomb branch are included in the captions of figures. The simple root fugacities used in the monopole formula computation of the Hilbert series are shown inside the quiver nodes or in a separate figure. The rest of the analysis aims to provide evidence for Conjecture \eref{Main Conjecture}.\\

 First, we state the anticipation of the global symmetry ($G_{global}$) on the Coulomb branch based on a conjectured claim about the $G_{global}$ of unbalanced quivers. We present the claim for \emph{minimally unbalanced} quivers in the beginning of section \ref{1}. Section \ref{3} contains an extended version of the claim for quivers with more than one unbalanced node. The analysis then further proceeds by the following steps:
 
\begin{itemize}
\item After a computation of the Hilbert series using simple root fugacities the unrefined Hilbert series (HS), obtained by setting all root fugacities to unity, is computed. The result is reported together with the corresponding expansion of the unrefined HS.
\item The Plethystic Logarithm (PL) of the unrefined HS is taken\footnote{For the definition of Plethystic Logarithm, see \cite{PL07} or (4.2) in \cite{HM11}.}. The $t^2$ coefficient is compared with the dimension of the adjoint representation of the expected $G_{global}$. This provides a necessary confirmation that the anticipated $G_{global}$ is correct. In cases of Coulomb branches which have a free sector (this happens for quivers containing a node with a negative imbalance) the global symmetry has two parts:
\begin{itemize}
\item Firstly, the freely generated part of the Coulomb branch is determined.
\item Secondly, this free sector is factored out so that the non-trivial part of the Coulomb branch can be further analyzed.
\end{itemize}
\item The fugacity map, that turns the simple root fugacities into the appropriate fugacities of the $G_{global}$ is given. One then shows that the $t^2$ coefficient of the refined HS is the character of the adjoint representation of $G_{global}$. This serves as a direct verification of the $G_{global}$ of the theory\footnote{In case of quivers with large character coefficients, this step is by-passed by showing directly the refined PL in the next step.}. (In case of a theory which has a free sector, the free sector appears in the form of a character coefficient in front of $t$ in the refined Hilbert series. Before proceeding further with the refined analysis, the free sector is factored out by multiplying the refined HS with an inverse of the Plethystic Exponential\footnote{For details of Plethystic Exponential see \cite{PL07} or (4.2) in \cite{HM11}.} (PE) of the character appearing in front of the $t$ term.) 
\item Next, the Plethystic Logarithm (PL) of the refined HS is taken. The refined PL encodes the information about the number, degree and representation behavior of generators and relations which define the Coulomb branch as an affine algebraic variety. 
\item The representation content of the chiral ring can be described using a simple polynomial. This compact form is given by the Highest Weight Generating function (HWG) \cite{HWG}. The HWGs for the first two families of quivers have simple forms and are therefore included. All HWGs are given in the form of Plethystic Exponential (PE). 
\item Finally, the Coulomb branch is identified.
\end{itemize}
After the exhaustion of all partitions for a given parent quiver (i.e. when all quivers obtainable by discrete gauging on the parent quiver are exhausted), the volumes of the Coulomb branches are compared. The corresponding ratios are summarized in tables at the end of each subsection. This serves as a non-trivial check that the Coulomb branches of the constructed daughter quivers are orbifolds of the parent Coulomb branch. Section \ref{2} contains a natural generalization of the results of section \ref{1} to theories with $SU(2)^{n_1}\times D_{n_2+1}$ global symmetry and an identification of a general pattern of HWG for the \emph{second family} of quivers. In section \ref{3} the \emph{third family} of theories is studied. In particular, we examine the $k=n=3$ representative of $A$-type bouquet quivers invariant under an additional $\mathbb{Z}_2$ outer automorphism symmetry. Section \ref{4} discusses discrete gauging on Coulomb branches of non-simply laced theories. A particular example of a bouquet quiver with a factor of $C_2$ in the global symmetry is studied. We conclude and discuss possible directions of further investigations in section \ref{5}, where we also propose a broader generalization of the pattern of HWG formula \eref{eq:Gen} found at the end of section \ref{2}.


\section{First Family: Quivers with central $2$ node and a bouquet of $1$ nodes} \label{1}

In order to begin, consider the theory in Figure \eref{fig:4} and set $k=2$. For $k=2$ the discrete global symmetry of the bouquet enhances from $S_n$ to $S_{n+2}$ (i.e. the bouquet of $n$ rank $1$ nodes enhances to a bouquet of $n+2$ nodes). One obtains the simplest quivers for the study of discrete gauging and orbifold actions.  

\subsection{Case: $k=2$, $n=1$}

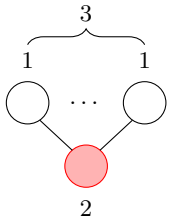
\begin{figure}[h!]
\center{
\begin{tikzpicture}[scale=0.70]
\draw (0,0)[red,fill=red!30] circle (0.4cm);
\draw (0,-0.8) node {\footnotesize{$2$}};
\draw (-0.26,0.3) -- (-0.87,0.87);
\draw (-1.1,1.2) circle (0.4cm);
\draw (-1.1,2) node {\footnotesize{$1$}};
\draw (0.26,0.3) -- (0.87,0.87);
\draw (1.1,1.2) circle (0.4cm);
\draw (1.1,2) node {\footnotesize{$1$}};
\draw (0,1.2) node {\footnotesize{$\dots$}};
\draw [decorate,decoration={brace,amplitude=6pt}] (-1.1,2.3) to (1.1,2.3);
\draw (0,2.9) node {\footnotesize{$3$}};
\end{tikzpicture}
}
\caption{\label{fig:D4-01} $\mathcal{P}_{[1^3]}(3)$ Quiver with $SU(2)^3 \subset Sp(4)$ global symmetry, $b=-1$, $dim^\mathbb{H} \mathcal{M}_C  = 4$.}
\end{figure}

Further, lets consider the $n=1$ case. The trivial $S_1$ symmetry enhances to $S_3$ discrete global symmetry, which becomes the group of outer automorphisms of the quiver permuting the bouquet nodes. Correspondingly, the theory is denoted by $\mathcal{P}_{[1^3]}(3)$. The quiver forms the finite $D_4$ Dynkin diagram depicted in Figure \eref{fig:D4-01}, which is the only Dynkin diagram with the the \emph{triality} property. 

For a simply laced quiver the balance of the $i$-th node is defined as \cite{HK16}:
\be
b_{ADE}(i) = \sum_{j\in \: adjacent \: nodes} N_j - 2N_i ,
\ee
where $N$ denotes the rank. Quivers with a single unbalanced node (i.e. single node with balance $b \neq 0$) are termed \emph{minimally unbalanced}. Throughout this paper the unbalanced nodes are conveniently drawn red. The red node of the minimally unbalanced quiver in Figure \eref{fig:D4-01} has balance $b=3\times1 -2\times2=-1$. Negative balance indicates that the theory has a free sector, which implies that either part of, or the entire Coulomb branch, is \emph{freely generated} \footnote{See observation 3.1 in \cite{FG08}.}. The identification of the global symmetry of a minimally unbalanced quiver is based on the following important claim\footnote{Extended version of this claim, applicable for quivers with two or more unbalanced nodes, is formulated in section \ref{3}.}:\\

\emph{Given a minimally unbalanced quiver $\sf Q$, the global symmetry on the Coulomb branch is:
$G_{global} =\prod_{i} G_i $,
where $G_i$ are groups corresponding to the Dynkin diagrams that are formed by the subsets of balanced nodes of $\sf Q$.}\\

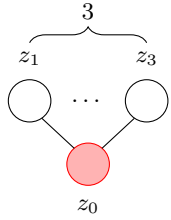
\begin{figure}[h!]
\center{
\begin{tikzpicture}[scale=0.70]
\draw (0,0)[red,fill=red!30] circle (0.4cm);
\draw (0,-0.8) node {\footnotesize{$z_0$}};
\draw (-0.26,0.3) -- (-0.87,0.87);
\draw (-1.1,1.2) circle (0.4cm);
\draw (-1.1,2) node {\footnotesize{$z_1$}};
\draw (0.26,0.3) -- (0.87,0.87);
\draw (1.1,1.2) circle (0.4cm);
\draw (1.1,2) node {\footnotesize{$z_3$}};
\draw (0,1.2) node {\footnotesize{$\dots$}};
\draw [decorate,decoration={brace,amplitude=6pt}] (-1.1,2.3) to (1.1,2.3);
\draw (0,2.9) node {\footnotesize{$3$}};
\end{tikzpicture}
}
\caption{\label{fig:D4-01FUG} $\mathcal{P}_{[1^3]}(3)$ Quiver with fugacities.}
\end{figure}

Since the balanced sub-quivers in Figure \eref{fig:D4-01} correspond to three $A_1$ Dynkin diagrams, the global symmetry is expected to be $SU(2) \times SU(2) \times SU(2)$. Moreover, each of the three bouquet nodes that connects to the unbalanced node contributes with a fundamental representation of $SU(2)$. As a consequence, there are $8$ monopole operators transforming under the three-fundamental representation of $SU(2)^3$, denoted by Dynkin labels $[1;1;1]$. These monopole operators carry spin 1/2 charge under $SU(2)_R$ (i.e. the R-symmetry). As 8 is also the complex dimension of the Coulomb branch, we learn that the whole Coulomb branch is free and it is a copy of $\qu^4$ with a global symmetry $Sp(4)$. Hence, for the global symmetry we can write:
\be
G_{global}=SU(2) \times SU(2) \times SU(2) \subset Sp(4),
\ee
where the explicit embedding is given in Equation \eref{eq:emb1}
\be \label{eq:emb1}
[1;1;1]_{SU(2) \times SU(2) \times SU(2)} \hookleftarrow [1,0,0,0]_{Sp(4)}.
\ee 
In order to find the global symmetry explicitly, one computes the Hilbert Series, utilizing the monopole formula \cite{CHZ14}. One first starts with the assignment of simple root fugacities given in Figure \eref{fig:D4-01FUG}.
 As is outlined in the introduction,
one then proceeds by computing the unrefined Hilbert series (HS), which is obtained by setting all simple root fugacities to unity: $z_i = 1, i=0,1,2,3$. The unrefined Hilbert series is given by Equation \eref{eq:hs01}.
\begin{equation} \label{eq:hs01}
HS(t)=\frac{1}{(1-t)^8}
\end{equation}
The expansion of the unrefined Hilbert series reads
\be \label{eq:uHS1}
HS(t) =1+8t +36t^2 +120t^3+330 t^4 + 792 t^5 + O(t^6).
\ee
Taking the Plethystic Logarithm (PL) of the unrefined Hilbert series one finds
\be
PL= 8t.
\ee
The following can be immediately observed:
\begin{itemize}
\item The absence of any negative contributions (absence of relations) signifies that the entire Coulomb branch is freely generated. 
\item The $t$ coefficient corresponds to dimension of some representation of the $G_{global}$.
\end{itemize}
Based on the previous claim, the $t$ coefficient corresponds to the dimension of the three-fundamental representation (rep) of $SU(2)^3$
\be
dim [1;1;1] = 2 \times 2 \times 2 = 8,
\ee
where $[a_1;a_2;a_3]$ denote the Dynkin labels of the three-representation of $SU(2) \times SU(2) \times SU(2)$. Next step is to employ the fugacity map. The simple root fugacities $z_i$, $i=1,2,3$ are mapped to the $SU(2)$ fundamental weight fugacities $x_i$, $i=1,2,3$ according to prescription \eref{eq:fm-111} 
\begin{gather}
 z_i \rightarrow x_i^2,\quad i=1,2,3 \label{eq:fm-111} \\ 
  z_0 \rightarrow (z_1 z_2 z_3)^{-\frac{1}{2}}, \label{eq:fm-112}
 \end{gather}
and the simple root fugacity of the unbalanced node, $z_0$, is eliminated according to substitution \eref{eq:fm-112}. In case of a quiver with only gauge nodes, the mapping that eliminates the fugacity of an unbalanced node is canonically derived from the gauge fixing condition in the following way. Consider a minimally unbalanced quiver with simple root fugacities $z_i, i = 1, \dots, N$ and the corresponding node ranks $r_i, i = 1, \dots, N$. Without the loss of generality, let $z_N$ be the fugacity of the unbalanced node and $r_N$ its rank, respectively. Then, the elimination of $z_N$ is derived from the constraint:
\be \label{eq:constraint1}
\prod_{i}^{N} {z_i}^{r_i} = 1  \implies z_N = (\prod_{i}^{N-1} {z_i}^{r_i})^{-\frac{1}{r_N}}.
\ee
After the mapping, given by \eref{eq:fm-111} and \eref{eq:fm-112}, the expansion of the refined HS is computed as
\be \label{eq:rHS-1}
HS(x_i,t) = 1+ \left( \frac{1}{x_1 x_2 x_3} + \frac{x_1}{x_2 x_3} + \frac{x_2}{x_1 x_3} + \frac{x_1 x_2}{x_3} + \frac{x_3}{x_1 x_2} + \frac{ x_1 x_3}{x_2} + \frac{x_2 x_3}{x_1} + x_1 x_2 x_3\right)t 
+ O(t^2) .
\ee
Rewriting the $t$ coefficient as
\be
\left(x_1 + \frac{1}{x_1}\right)\left( x_2 + \frac{1}{x_2}\right)\left( x_3 +\frac{1}{x_3}\right)
\ee
one directly identifies the character of the three-fundamental representation od $SU(2)^3$, which verifies the expectation of the global symmetry. The expression of the refined HS is compactly written in Equation \eref{eq:R001}
\begin{equation} \label{eq:R001}
HS(x_i,t)=PE[[1;1;1]t]=\prod_{\epsilon_i =\pm 1} \frac{1}{1-x_1 ^{\epsilon_1} x_2 ^{\epsilon_2} x_3 ^{\epsilon_3} t} ,
\end{equation}
where $x_i, \;i=1,2,3$ are the fugacities of the fundamental weights of $SU(2)^3$ and $\epsilon_i$ runs over the two weights of the fundamental representation of $SU(2)$. Equation \eref{eq:R001} coincides with (4.1) in \cite{HM11}. In general, as an (affine) algebraic variety, the Coulomb branch is specified by:
\begin{itemize}
\item Number and degree of generators 
\item Representation under which generators transform (to all relevant orders of $t$)
\item Representations under which relations transform (to all relevant orders of $t$)
\end{itemize}
All this information is succinctly encoded in the Plethystic Logaritm (PL) of the refined Hilbert series. Taking the PL of the refined Hilbert series in Equation \eref{eq:rHS-1} or \eref{eq:R001} one obtains Equation (\ref{eq:PL1}),
\begin{equation} \label{eq:PL1}
PL= [1;1;1]_8t
\end{equation}
where the subscript denotes the total complex dimension of the representation $dim^{\mathbb{C}} [1;1;1] = 2 \times 2 \times 2 = 8$. The subscript notation of the refined PL is conveniently used throughout the paper to denote the dimensions of the corresponding representations. The Coulomb branch is a freely generated space of quaternionic dimension $4$:
 \be
 \mathcal{C}_{[1^3]} = \mathbb{H}^4 .
 \ee
The representational content of the chiral ring is neatly encoded by the highest weight generating function (HWG). The HWG for the theory in Figure \eref{fig:D4-01} is given by Equation \eref{eq:hwg001}, which agrees with (4.3) in \cite{HM11},
\be \label{eq:hwg001}
HWG=PE[\mu_1 \mu_2 \mu_3 t + \sum_{i=1}^3 \mu_i^2 t^2 + \mu_1 \mu_2 \mu_3 t^3 +t^4 -\mu_1^2 \mu_2^2 \mu_3^2 t^6]
\ee
and where $\mu_i,\; i=1,2,3$ are the highest weight fugacities of the three $SU(2)$ representations.
In terms of $Sp(4)$ representations, the HWG takes the simple form
\be
HWG = PE \left [ \mu_1 t \right ],
\ee
where now $\mu_1$ denotes a highest weight fugacity for $Sp(4)$.

\subsubsection{Gauging $H_{\lambda} = \mathbb{Z}_2$}

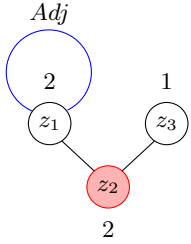
\begin{figure}[h!]
\center{
\begin{tikzpicture}[scale=0.70]
\draw (0,0)[red,fill=red!30]  circle (0.4cm);
\draw (0,-0.8) node {\footnotesize{$2$}};
\draw (0,0) node {\footnotesize{$z_2$}};
\draw (-0.26,0.3) -- (-0.87,0.87);
\draw (-1.1,1.2) circle (0.4cm);
\draw (-1.1,2) node {\footnotesize{$2$}};
\draw (-1.1,1.2) node {\footnotesize{$z_1$}};
\draw (0.26,0.3) -- (0.87,0.87);
\draw (1.1,1.2) circle (0.4cm);
\draw (1.1,2) node {\footnotesize{$1$}};
\draw (1.1,1.2) node {\footnotesize{$z_3$}};
 \draw[-,blue] (-0.765,1.455) arc (-64:247:0.8);
 \draw (-1.1,3.27) node {\footnotesize{$Adj$}};
\end{tikzpicture}
}
\caption{\label{fig:D4-02} $\mathcal{P}_{[2,1]}(3)$ Quiver with $SU(2)^2 \cong SO(4)$ global symmetry, $b=-1$, $dim^\mathbb{H} \mathcal{M}_C  = 4$.}
\end{figure}

Next, we would like to construct a new theory with a Coulomb branch that is an orbifold of the Coulomb branch of the previously analyzed $\mathcal{P}_{[1^3]}(3)$ theory. Assuming Conjecture \eref{Main Conjecture}, let us gauge a subgroup $\mathbb{Z}_2 \subset S_3$ of the discrete global symmetry of the $\mathcal{P}_{[1^3]}(3)$ theory, which acts on the bouquet by permuting its three nodes. Following Conjecture \eref{Main Conjecture}, the bouquet of the constructed quiver consists of a single rank $1$ node and a single adjoint $2$ node. Accordingly, lets denote the newly constructed theory by $\mathcal{P}_{[2,1]}(3)$. The quiver and the explicit assignment of the simple root fugacities are depicted in Figure \eref{fig:D4-02}. Note that the adjoint $2$ node connected to a rank $2$ node is balanced. More generally, any adjoint node, with rank $N$, connected to a rank $2$ node is balanced because the extra hypermultiplet contributions coming from the adjoint loop exactly cancel the contributions from the vector multiplet.
Since there are two balanced $A_1$ sub-quivers, the expected global symmetry is $SU(2) \times SU(2) \simeq SO(4)$. After the computation of the Hilbert series using the simple root fugacities, set $\forall i,\;z_i =1$ to obtain the unrefined HS in Equation \eref{eq:uHS02}.
\be \label{eq:uHS02}
HS(t)=\frac{1+t^2}{(1-t)^6 (1-t^2)^2}
\ee
Equation \eref{eq:uHS02} has the expansion of the form
\be
HS(t)=1 + 6 t + 24 t^2 + 74 t^3 + 194 t^4 +O(t^5).
\ee
Taking the PL of the unrefined HS one obtains Equation \eref{eq:uPL1}
\be \label{eq:uPL1}
PL=6 t + 3 t^2 - t^4.
\ee
The term by term analysis of Equation \eref{eq:uPL1} implies the following:
\begin{itemize}
\item $t$: there is a freely generated part of the Coulomb branch with quaternionic dimension $3$. Thus, the free part of the Coulomb branch is \cite{MOTZ17}: $\mathcal{C}_{f.g.} = \mathbb{H}^3$, which is generated by the fundamental representation of $Sp(3)$, denoted by Dynkin labels $[1,0,0]$. Hence, the global symmetry has two constituent parts:
\be
G_{global} = G_{global, free} \times G_{global,\; non-trivial},
\ee
such that $G_{global, free} = Sp(3)$. 
\item $t^2$: there is a non-trivial part of the Coulomb branch generated by a $3$ dimensional representation of $G_{global, \;non-trivial}$. In order to analyze the non-trivial part of the Coulomb branch, the free part needs to be multiplied out.
\item $t^4$: there is a relation at this order that transforms as a singlet under $G_{global}$.
\end{itemize}
Utilize the fugacity map
\begin{gather}
 z_1 \rightarrow x_1^2, \label{eq:E01} \\
  z_3 \rightarrow x_2^2, \label{eq:E02} \\
  z_2 \rightarrow (z_1^2 z_3)^{-\frac{1}{2}}, \label{eq:E1}
 \end{gather}
 where the simple root fugacities $z_1, z_3$ map to $x_1$ and $x_2$, the fundamental weight fugacities of the two $SU(2)$ following prescriptions \eref{eq:E01} and \eref{eq:E02}, respectively. Note, that the latter $SU(2)$ corresponds to the rank $1$ bouquet node and the former $SU(2)$ corresponds to the adjoint $2$ node. The $z_2$ root fugacity of the unbalanced node is eliminated according to prescription \eref{eq:E1}, which again, follows from constraint \eref{eq:constraint1}. The expansion of the refined HS takes the form
 \be \label{eq:euHS2}
 \begin{split}
 HS(x_1,x_2,t)&= 1+ \left(x_1^2 + 1 + \frac{1}{x_1^2}\right)\left(x_2 + \frac{1}{x_2}\right)t +\left(x_2^2 +1 + \frac{1}{x_2^2}\right)t^2\\
 &+\left( x_2^4 +x_2^2 +1 + \frac{1}{x_2^2} +\frac{1}{x_2^4} \right)t^4 + O(t^6)
 \end{split}
 \ee
Comparing Equations \eref{eq:euHS2} and \eref{eq:uPL1} one infers that the $t$ coefficient in Equation \eref{eq:uPL1} can be regarded as the dimension of the $[2;1]$ two-representation of $SU(2)\times SU(2)$ (or the dimension of the $[1,0,0]$ fundamental representation of $Sp(3)$). In this case one finds the embedding\footnote{We write $SO(3)$ instead of $SU(2)$ given the universal double covering of the $A_1$ algebra and since the $[2]$ representation is real.}:
\be 
[2;1]_{SO(3) \times Sp(1)} \hookleftarrow [1,0,0]_{Sp(3)}.
\ee
Observe in Equation \eref{eq:euHS2} that the $SU(2)$ symmetry of the adjoint $2$ node appears in the symmetry of the free sector at order $t$ but at higher orders of $t$ only the $SU(2)$ that corresponds to the balanced $1$ node plays a role. Let us now multiply out the free sector of the theory and continue with the analysis of the non-trivial part of the Hilbert Series. One takes the PE of the character in front of the $t$ coefficient and multiplies the whole HS by the inverse of this PE. The obtained refined HS now describes the non-trivial part of the Coulomb branch. The corresponding refined PL can be written as:
\be \label{eq:pleuHS2}
PL= [2;0]_{3}t^2 - [0;0]_1t^4.
\ee
The $t^2$ coefficient can be regarded as the $3$ dimensional representation of $SU(2) \times SU(2)$ with Dynkin labels $[2;0]$. The relation at $t^4$ transforms as a singlet under the two-representation of $SU(2)\times SU(2)$, denoted by $[0;0]$. Note that the simple PL in Equation \eref{eq:pleuHS2} describes a $\mathbb{C}^2/\mathbb{Z}_2$ singularity, also termed $A_1$ singularity in the literature. In summary, the Coulomb branch of the theory:
\be
\mathcal{C}_{[2,1]} = \mathbb{H}^3 \times \mathbb{C}^2 / \mathbb{Z}_2 
\ee
has two parts; a free sector in the form of $\mathbb{H}^3$ and a non-trivial part in the form of an $A_1$ singularity \cite{CH16}. The representation content of the chiral ring on the non-trivial part of the Coulomb branch is described by the HWG in Equation \eref{eq:hwg11},
\be \label{eq:hwg11}
HWG=PE[\mu^2 t^2]
\ee
where $\mu$ denotes the highest weight fugacity of $SU(2)$.

\subsubsection{Gauging $H_{\lambda} = S_3$}

\begin{figure}[h!]
\center{
\begin{tikzpicture}[scale=0.70]
\draw (0,0)[red,fill=red!30]  circle (0.4cm);
\draw (0,-0.8)node {\footnotesize{$2$}};
\draw (0,0)node {\footnotesize{$z_2$}};
\draw (0,0.4) -- (0,0.8);
\draw (0,1.2) circle (0.4cm);
\draw (0,2) node {\footnotesize{$3$}};
\draw (0,1.2) node {\footnotesize{$z_1$}};
 \draw[-,blue] (0.335,1.455) arc (-64:247:0.8);
 \draw (0,3.27) node {\footnotesize{$Adj$}};
\end{tikzpicture}
}
\caption{\label{fig:D4-03} $\mathcal{P}_{[3]}(3)$ Quiver with $SU(2) \subset Sp(2)$ global symmetry, $b=-1$, $dim\; \mathcal{M}_C ^\mathbb{H} = 4$.}
\end{figure}
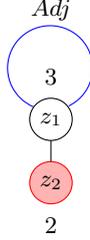

We saw that the discrete gauging of $H_{\lambda}=\mathbb{Z}_2 \subset S_3$ for the parent $\mathcal{P}_{[1^3]}(3)$ quiver produced the $\mathcal{P}_{[2,1]}(3)$ quiver theory with a Coulomb branch in the form of a discrete quotient:
\be
\mathcal{C}_{[2,1]} = \mathcal{C}_{[1^3]}/\mathbb{Z}_2.
\ee 
Given the success of this construction, let us consider the $\mathcal{P}_{[1^3]}(3)$ theory where the entire discrete $S_3$ global symmetry is gauged. According to Conjecture \eref{Main Conjecture}, this theory is described by the quiver, depicted with explicit assignment of the simple root fugacities, in Figure \eref{fig:D4-03}. The quiver consists of a single adjoint $3$ node attached to the central $2$ node. Accordingly, the theory is denoted by $\mathcal{P}_{[3]}(3)$.
The balanced part of the quiver forms the $A_1$ Dynkin diagram, therefore, the anticipated global symmetry is $SU(2)$. The unrefined HS, obtained by setting all the simple root fugacities $z_i$ to unity, takes the form given by Equation \eref{eq:uHS3}
 \be \label{eq:uHS3}
HS(t)=\frac{1 + t^2 + 2 t^3 + t^4 + t^6}{(1 - t)^4 (1 - t^2)^2 (1 - t^3)^2}.
\ee
The expansion of Equation \eref{eq:uHS3} yields
\be
HS(t)=1 + 4 t + 13 t^2 + 36 t^3 + 87 t^4 + 190 t^5 + 386 t^6 +O(t^7).
\ee
The PL of the unrefined expression \eref{eq:uHS3} reads
\be \label{eq:typo1}
PL=4 t + 3 t^2 + 4 t^3 - 2 t^5 -3t^6+ O(t^7).
\ee
Let us analyze the first two terms in the last expression:
\begin{itemize}
\item $t$: there is a free sector corresponding to $\mathbb{H}^2$ generated by $G_{global,\;free} = Sp(2)$. In particular, it is generated by the fundamental representation of $Sp(2)$, denoted by Dynkin labels $[1,0]$.
\item $t^2$: the coefficient matches the dimension of the adjoint representation of $SU(2)$. Hence, the expected non-trivial global symmetry is: $G_{global,\; non-trivial}=SU(2)$.
\end{itemize}
 The refined expression for the Hilbert series is obtained using the fugacity map
 \begin{gather}
 z_1 \rightarrow x^2, \\
  z_2 \rightarrow (z_1^3)^{-\frac{1}{2}}, \label{eq:E2}
 \end{gather}
where the simple root fugacity $z_1$ maps to the $SU(2)$ fundamental weight fugacity $x$, and the root fugacity of the unbalanced node, $z_2$, is eliminated according to the gauge fixing condition \eref{eq:E2}. The computation of the refined HS yields Equation \eref{eq:rHS03}.
\be \label{eq:rHS03}
\begin{split}
HS_{[3]}(t,x)&=1 + \left(\frac{1}{x^3} + \frac{1}{x}  + x + x^3 \right)t+ \left(\frac{1}{x^6}+\frac{1}{x^4}+\frac{3}{x^2}+3x^6 +3x^2+ x^4+ x^6\right) t^2 \\
&+ O(t^3).
\end{split}
 \ee
We see that the free sector, corresponding to $\mathbb{H}^2$, is spanned by generators in the fundamental rep of $Sp(2)$, denoted by $[1,0]$, or equivalently in the $[3]_4$ rep of $SU(2)$. The embedding found in this case can be written as:
\be
 [3]_{SU(2)} \hookleftarrow [1,0]_{Sp(2)} .
 \ee
 Let us multiply out the free sector in an analogous manner as in the previous case. Again, one multiplies the refined HS by the inverse of the PE of the character in front of $t$ in Equation \eref{eq:rHS03}. The PL of the obtained refined HS for the non-trivial part of the Coulomb branch takes the form:
 \be
 \begin{split}
 PL&= \left(\frac{1}{x^2}+1 +x^2\right) t^2 + \left(\frac{1}{x^3} + \frac{1}{x}  + x + x^3\right)t^3 \\
&- \left(\frac{1}{x} + x \right)t^5 - \left(\frac{1}{x^2}+1 +x^2\right) t^6+ O(t^7).
\end{split}
\ee
The character appearing in front of $t^2$ is the character of the adjoint representation of the non-trivial global symmetry $SU(2)$. This verifies that
\be
G_{global,\;non-trivial}=SU(2).
\ee
Now, the refined PL can be written as
\be\label{eq:pl13}
PL= [2]_3 t^2 +[3]_4t^3- [1]_2 t^5 -[2]_3t^6 +O(t^7),
\ee
where $[a]$ denotes the Dynkin labels of the representation of $SU(2)$. Comparing this expression with the refined HS:
\be \label{eq:hhh1}
HS=1+[2]t^2+[3]t^3+([4]+[0])t^4+([5]+[3])t^5+(2[6]+[2])t^6 + O(t^7)
\ee
one sees that at order $t^5$ there is one operator that must be set to zero, and at order $t^6$ there are two operators that satisfy one relation, hence must be proportional to each other. These observations are used bellow in the explicit construction of the Coulomb branch algebraic variety. The HWG for the non-trivially generated part of the Coulomb branch is given by Equation \eref{eq:hwg13}
\begin{equation} \label{eq:hwg13}
HWG= PE[\mu^2 t^2 + \mu^3 t^3 + t^4 + \mu^3 t^5  -\mu^6 t^{10}]
\end{equation}
where $\mu$ is the highest weight fugacity of $SU(2)$.\\
 
The $\mathcal{P}_{[3]}(3)$ theory has none of the $S_3$ global symmetry compared to the $\mathcal{P}_{[1^3]}(3)$ theory. Therefore, by Conjecture \eref{Main Conjecture} it is implied that the $\mathcal{C}_{[3]}$ Coulomb branch is an $S_3$ orbifold of the parent $\mathcal{C}_{[1^3]}$ Coulomb branch. Indeed, in accord with Equation \eref{eq:00}, the computed Coulomb branch variety can be written as:
\be \label{eq:coul1}
\mathcal{C}_{[3]} = \mathbb{H}^2 \times \mathbb{C}^4 / S_3.
\ee
This result is confirmed by explicit computations of the $S_3$ Molien invariant reproducing Equation \eref{eq:uHS3}.

Let us analyze Equation \eref{eq:pl13} and \eref{eq:hhh1} in more detail. In \eref{eq:pl13} there are generators transforming under the adjoint $[2]$ rep of $SU(2)$ at $t^2$, and additional generators at order $t^3$ transforming under the $[3]$ rep. Altogether we have $7$ generators. There are relations at order $t^5$ and $t^6$ transforming under $[1]$ and $[2]$ reps, respectively. Explicitly, the generators at $t^2$ are:
 \be
 M_{\alpha \beta},
 \ee
where $\alpha, \beta = 1,2$, and they satisfy 
 \begin{gather}
  M_{\alpha \beta} =  M_{ \beta \alpha}, \\
  deg(M) = 2,
  \end{gather}
  where $deg()$ denotes the degree of the generator which is associated with the power of $t$ at which they appear. The generators at $t^3$ are:
  \be
   N_{\alpha \beta \gamma},
   \ee
  where $\alpha, \beta, \gamma = 1,2$. These are also symmetric in all indices and with $deg(N)=3$. 
Now, remembering the tensor products
  \begin{gather}
  [2]\otimes [3] = [5]\oplus [3] \oplus [1]  \\
  Sym^2[3]_{10} = [6]_7 \oplus[2]_3 \label{eq:jj1}\\
  Sym^3[2]_{10}=[6]_7 \oplus[2]_3 \label{eq:jj2}
  \end{gather}
  and observing that in expression \eref{eq:hhh1} at order $t^5$ the $[1]$ is missing, one deduces that this must be a relation. Hence, the relation at order $t^5$ with degree $5$ is:
  \be  \label{eq:RR1}
  {\epsilon}^{\beta \delta}  {\epsilon}^{\alpha \gamma}  M_{\alpha \beta}  N_{\gamma \delta \epsilon} = 0.
  \ee
 At $t^6$ the two operators of degree $6$ satisfy Equation \eref{eq:RR2}.
 \be \label{eq:RR2}
   M_{\alpha_1 \alpha_2} M_{\alpha_3 \alpha_4} M_{\alpha_5 \alpha_6} {\epsilon}^{\alpha_2 \alpha_3} {\epsilon}^{\alpha_4 \alpha_5} = N_{\alpha_1 \alpha_2 \alpha_3}N_{\alpha_4 \alpha_5 \alpha_6}{\epsilon}^{\alpha_2 \alpha_4}{\epsilon}^{\alpha_3 \alpha_5}
   \ee  
Note that on the left hand side the operator transforms in the $[2]$ of \eref{eq:jj2} that is coming from the third symmetrization. The operator on the right hand side transforms in the $[2]$ in \eref{eq:jj1} which comes from the second symmetrization. Equations \eref{eq:RR1} and \eref{eq:RR2} produce $2+3=5$ equations that constrain the $7$ generators. The Coulomb branch can be computed from this explicit analysis employing Maclauay2\footnote{Maclauay2 program for computation of algebraic varieties is available at https://faculty.math.illinois.edu/Macaulay2/.}. The computation yields an unrefined HS of the form:
 \be
 HS(t)=\frac{1 - 2t^5 - 3 t^6 + 3t^8 +2t^9- t^{14}}{ (1 - t^2)^3 (1 -t^3)^4}
 \ee
 which is precisely the unrefined HS obtained previously if one factors out the free sector in Equation \eref{eq:uHS3}. Let us now use the comparison of the Coulomb branch volumes as a non-trivial test of Conjecture \eref{Main Conjecture}. \\

\subsubsection{Comparison of the Coulomb branch volumes}

\begin{figure}[h!]
\center{
\begin{tikzcd}[row sep=2.5em]
& \mathcal{C}_{[1^3]} \arrow{dd}{S_3} \arrow{dl}{\mathbb{Z}_2} \\
\mathcal{C}_{[2,1]} \arrow{dr}{\mathbb{Z}_3} & 
\\ & \mathcal{C}_{[3]}
\end{tikzcd}
}
\caption{\label{fig:Commut2} Commutative Diagram of Coulomb branch orbifolding for $k=2$, $n=1$ theories.}
\end{figure}
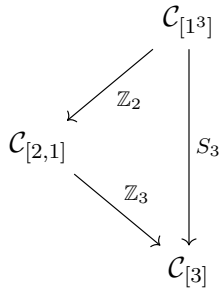

Following the method outlined in section \ref{0} for the $k=2$, $n=1$ theories, one can compare the volumes of the Coulomb branches. Expanding the unrefined Hilbert series \eref{eq:uHS1}, \eref{eq:uHS02} and \eref{eq:uHS3} according to Equation \eref{eq:expansionHS} and plugging into Equation \eref{eq:01} one finds:
 \begin{gather}
 \frac{vol(\CC_{[1^3]})}{vol(\CC_{[2,1]})} = \frac{R_{[1^3]}}{R_{[2,1]}} = \frac{1} {\frac{1}{2}} = 2 = ord(\mathbb{Z}_2) \label{eq:R01}  \\
  \frac{vol(\CC_{[1^3]})}{vol(\CC_{[3]})} = \frac{R_{[1^3]}}{R_{[3]}} = \frac{1} {\frac{1}{6}} = 6 = ord(S_3) \label{eq:R02} 
 \end{gather}
 which are the expected ratios. Equations \eref{eq:R01} and \eref{eq:R02} provide a non-trivial test of Conjecture \eref{Main Conjecture}, namely, that the Coulomb branches of $\mathcal{P}_{[2,1]}(3)$ and $\mathcal{P}_{[3]}(3)$ are $\mathbb{Z}_2$ and $S_3$ orbifolds of the parent $\mathcal{P}_{[1^3]}(3)$ Coulomb branch, respectively. Note, that it follows that the Coulomb branch of $\mathcal{P}_{[3]}(3)$ is a $\mathbb{Z}_3$ quotient of the $\mathcal{P}_{[2,1]}(3)$ Coulomb branch. This can be tested explicitly employing the ideas of stepwise projection \cite{Stepwise}. Table \eref{tab:Ratios01} summarizes the ratios of volumes between the Coulomb branches of $k=2$, $n=1$ theories.
\begin{table}
\begin{center}
\begin{tabular}{ |p{1.6cm}||p{1.5cm}|p{1.5cm}|p{1.5cm}|}
 \hline
 \multicolumn{4}{|c|}{Volume Ratios of $k=2$, $n=1$ theories} \\
 \hline
 Partition & $[1^3]$&$[2,1]$&$[3]$\\
 \hline \hline
 $[1^3]$&$1$&2&$6$\\
 $[2,1]$&&$1$&$3$\\
 $[3]$&&&$1$\\
 
 \hline
\end{tabular}
\end{center}
\caption{\label{tab:Ratios01} Ratios of Coulomb branch volumes for $k=2$, $n=1$ theories.}
\end{table}
The relations between the Coulomb branches of $k=2$, $n=1$ theories are schematically depicted by the commutative diagram in Figure \eref{fig:Commut2}, where the arrows denote quotients. 

\newpage

\subsection{Case: $k=2$, $n=2$} 

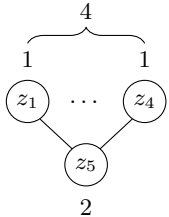
\begin{figure}[h!]
\center{
\begin{tikzpicture}[scale=0.70]
\draw (0,0) circle (0.4cm);
\draw (0,-0.8) node {\footnotesize{$2$}};
\draw (0,0) node {\footnotesize{$z_5$}};
\draw (-0.26,0.3) -- (-0.87,0.87);
\draw (-1.1,1.2) circle (0.4cm);
\draw (-1.1,2) node {\footnotesize{$1$}};
\draw (-1.1,1.2) node {\footnotesize{$z_1$}};
\draw (0.26,0.3) -- (0.87,0.87);
\draw (1.1,1.2) circle (0.4cm);
\draw (1.1,2) node {\footnotesize{$1$}};
\draw (1.1,1.2) node {\footnotesize{$z_4$}};
\draw (0,1.2) node {\footnotesize{$\dots$}};
\draw [decorate,decoration={brace,amplitude=6pt}] (-1.1,2.3) to (1.1,2.3);
\draw (0,2.9) node {\footnotesize{$4$}};
\end{tikzpicture}
}
\caption{\label{fig:D4-1} $\mathcal{P}_{[1^4]}(4)$ Quiver with $D_4$ global symmetry, $b=0$, $dim\; \mathcal{M}_C ^\mathbb{H} = 5$.}
\end{figure}

Let us now turn to the case $k= n =2$. The $S_2$ discrete global symmetry of the bouquet enhances to $S_4$. The five partitions of $4$ are $\mathcal{P}(4)=\{ [1^4], [2,1^2], [2^2,1], [3,1], [4] \}$. The first theory, corresponding to $\mathcal{P}_{[1^4]}(4)$ is shown in Figure \eref{fig:D4-1} together with the corresponding assignment of the simple root fugacities.
The quiver is fully balanced and forms the $\hat{D_4}$ Dynkin diagram (i.e. the affine Dynkin diagram of $D_4$). After the decoupling of the center of mass $U(1)$, one expects to find the enhanced $D_4 \equiv SO(8) \supset SU(2)^4$ global symmetry. The $SU(2)^4$ is the maximal subgroup of $SO(8)$ that has the natural $S_4$ symmetry which plays a role in the following analysis. The computation of the unrefined HS yields 
\begin{equation} \label{eq:hs21}
HS_{[1^4]}(t)=\frac{(1 + t^2) (1 + 17t^2 + 48t^4 + 17t^6 + t^8)}{(1-t^2)^{10}}
\end{equation}
Equation \eref{eq:hs21} is consistent with previous results in \cite{BHM10} and in Table 11 in \cite{HK16}. Expanding Equation \eref{eq:hs21}, one finds
\be
HS(t)=1 + 28 t^2 + 300 t^4 + 1925 t^6 + 8918 t^8 +O(t^{10}).
\ee
The corresponding PL reads
\be \label{eq:uPL21}
PL=28 t^2 - 106 t^4 + 833 t^6 - 8400 t^8 + O(t^{10})
\ee
The $t^2$ coefficient agrees with the dimension of the adjoint representation of $SO(8)$:
\be
dim\; [0,1,0,0]_{D_4} = 28.
\ee
There is a crucial difference between the PLs in the previous subsection and the PL in Equation \eref{eq:uPL21}. The absence of the $t$ term in Equation \eref{eq:uPL21} implies that there is no free sector (i.e. no free hypers) in the theory. This follows from the absence of a node with negative imbalance in the quiver. The simple root fugacities, indicated in Figure \eref{fig:D4-1} are treated in the following manner. As previously, one of the fugacities is eliminated by the gauge fixing condition. Recall, that the elimination condition follows from constraint \eref{eq:constraint1}. In this case, one eliminates one of the bouquet fugacities such that the remaining fugacities are in the shape of a $D_4$ Dynkin diagram. One declares the $z_4$ to be the null node (i.e. the affine node in the $\hat{D_4}$ Dynkin diagram) and the elimination of $z_4$ is thus based on prescription given by \eref{eq:E3}. Note that $z_4$ becomes the inverse of the adjoint weight fugacity. One uses the Cartan matrix of $D_4$ to map the remaining simple root fugacities $z_i,\; i=1,2,3,5$ to the fundamental weights of $D_4$, such that the powers in the fugacity map are determined by the components of the Cartan matrix. The mapping is summarized by the following equations:
 \begin{gather}
 z_4 \rightarrow (z_1 z_2 z_3 z_5^2)^{-1} = y_2^{-1}, \label{eq:E3} \\
 z_1 \rightarrow y_1^2 y_2^{-1}, \; z_2 \rightarrow y_2^2(y_1 y_3 y_4)^{-1}, \\
 z_3 \rightarrow y_3^2 y_2^{-1}, \; z_5 \rightarrow y_4^2 y_2^{-1}. 
 \end{gather}
 Making use of this fugacity map the refined HS is computed. One finds that the $t^2$ coefficient is precisely the character of the adjoint representation of $D_4$, which confirms that the global symmetry is $SO(8)$. For the purpose of brevity, we refrain from showing the character expansion of the refined Hilbert series and directly show the result of the computation of the refined PL:
\begin{multline}
PL=  [0,1,0,0]_{28}t^2 - ([0,0,0,0]_1+[2,0,0,0]_{35}+[0,0,2,0]_{35}+[0,0,0,2]_{35}) t^4 + \\
([2,0,0,0]_{35}+[0,1,0,0]_{28} + [0,0,2,0]_{35} + 2[1,0,1,1]_{350} + [0,0,0,2]_{35} )t^6 + O(t^8),
\end{multline}
where $[d_1,d_2,d_3,d_4]_{dim}$ are the Dynkin labels for $D_4$ and the subscript denotes the dimension of the representation. Note that the relations at order $t^4$ are manifestly invariant under the triality of $D_4$. The Coulomb branch is the reduced moduli space\footnote{In the literature, reduced single instanton moduli spaces are also known under the abbreviation RSIMS.} of one $D_4$-instanton on $\mathbb{C}^2$ \cite{BHM10,HK15}. Geometrically, the Coulomb branch is a simple algebraic variety which is a closure of the minimal nilpotent orbit\footnote{This space can be defined as a space of $8\times 8$ matrices $M$ that satisfy $M=-M^T,\; M^2=0,\;rank(M)\leq 2$.} of $D_4$:
\be
\mathcal{C}_{[1^4]}= \overline{min\mathcal{O}_{D_4}}.
\ee
All the information about the chiral ring is neatly encoded by the HWG in Equation \eref{eq:hwg1}
\begin{equation} \label{eq:hwg1}
HWG= PE[\mu_2 t^2],
\end{equation}
where $\mu_2$ is the fugacity of the highest weight of $D_4$.

\subsubsection{Gauging $H_{\lambda} = \mathbb{Z}_2$}

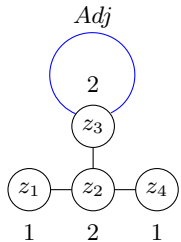
\begin{figure}[h!]
\center{
\begin{tikzpicture}[scale=0.70]
\draw (0,0) circle (0.4cm);
\draw (0,-0.8) node {\footnotesize{$2$}};
\draw (0,0) node {\footnotesize{$z_2$}};
\draw (0,0.4) -- (0,0.8);
\draw (0,1.2) circle (0.4cm);
\draw (0,2) node {\footnotesize{$2$}};
\draw (0,1.2) node {\footnotesize{$z_3$}};
\draw (-0.4,0) -- (-0.8,0);
\draw (-1.2,0) circle (0.4cm);
\draw (-1.2,-0.8) node {\footnotesize{$1$}};
\draw (-1.2,0) node {\footnotesize{$z_1$}};
\draw (0.4,0) -- (0.8,0);
\draw (1.2,0) circle (0.4cm);
\draw (1.2,-0.8) node {\footnotesize{$1$}};
\draw (1.2,0) node {\footnotesize{$z_4$}};
 \draw[-,blue] (0.335,1.455) arc (-64:247:0.8);
 \draw (0,3.27) node {\footnotesize{$Adj$}};
\end{tikzpicture}
}
\caption{\label{fig:D4-2} $\mathcal{P}_{[2,1^2]}(4)$ Quiver with $B_3$ global symmetry, $b=0$, $dim\; \mathcal{M}_C ^\mathbb{H} = 5$.}
\end{figure}

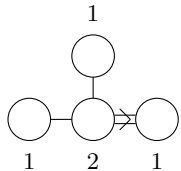
\begin{figure}[h!]
\center{
\begin{tikzpicture}[scale=0.70]
\draw (0,0) circle (0.4cm);
\draw (0,-0.8) node {\footnotesize{$2$}};
\draw (0,0.4) -- (0,0.8);
\draw (0,2) node {\footnotesize{$1$}};
\draw (-0.4,0) -- (-0.8,0);
\draw (-1.2,0) circle (0.4cm);
\draw (0,1.2) circle (0.4cm);
\draw (-1.2,-0.8) node {\footnotesize{$1$}};
\draw (0.4,0.08) -- (0.8,0.08);
\draw (0.4,-0.08) -- (0.8,-0.08);
\draw (0.5,0.2) -- (0.7,0);
\draw (0.5,-0.2) -- (0.7,0);
\draw (1.2,0) circle (0.4cm);
\draw (1.2,-0.8) node {\footnotesize{$1$}};
\end{tikzpicture}
}
\caption{\label{fig:b3no} Dynkin diagram of the affine $B_3$ algebra.}
\end{figure}

Given the evidence for Conjecture \eref{Main Conjecture} in the previous subsection for $k=2$, $n=1$ quivers, let us now use the $\mathcal{P}_{[1^4]}(4)$ quiver to construct theories for all other partitions of $\mathcal{P}(4)$. In order to construct the first theory, gauge $\mathbb{Z}_2 \subset S_4$, a subgroup of the discrete global symmetry of the parent $\mathcal{P}_{[1^4]}(4)$ quiver in Figure \eref{fig:D4-1}. According to Conjecture \eref{Main Conjecture}, the bouquet of the constructed theory consists of an adjoint $2$ node and two rank $1$ nodes. We denote this theory by $\mathcal{P}_{[2,1^2]}(4)$. The corresponding quiver and the simple root fugacities are depicted in Figure \eref{fig:D4-2}. Note that the adjoint node connected to rank $2$ node is balanced.
Since we are studying a quiver obtained by a $\mathbb{Z}_2$ quotient of a quiver with $SO(8)$ global symmetry and $SO(7)$ is a subgroup of $SO(8)$ that commutes with $\mathbb{Z}_2$, this provides the first indication for the expectation of the global symmetry. Another indication for the anticipated global symmetry comes from comparing the quiver in Figure \eref{fig:D4-2} with the affine $B_3$ Dynkin diagram, depicted in Figure \eref{fig:b3no}. 
When one eliminates one of the simply connected rank $1$ nodes (using the gauge fixing condition), it is natural to expect:
\be
G_{global} =B_3 \equiv SO(7).
\ee
Before we turn to the refined analysis, let us proceed by computing the HS using the simple root fugacities, and setting all to unity in order to obtain the expression of the unrefined HS:
\begin{equation} \label{eq:hs22}
HS_{[2,1^2]}(t)=\frac{(1 + t^2) (1 + 10t^2 + 20t^4 + 10t^6 + t^8)}{(1-t^2)^{10}}.
\end{equation} 
Indeed, note that Equation \eref{eq:hs22} contains the HS of the next to minimal nilpotent orbit of $B_3$, listed in Table 10 in \cite{HK16}. Expanding the unrefined HS, one obtains Equation \eref{eq:uPL2}
\be \label{eq:uPL2}
HS(t) = 1+ 21t^2 +195t^4 + 1155t^6 + 5096t^8 + O(t^{10}),
\ee
which has the PL of the form
\be
PL=21 t^2 - 36 t^4 + 140 t^6 - 784 t^8 + O(t^{10}).
\ee
The $t^2$ coefficient in the last expression is the dimension of the adjoint representation of $B_3 \equiv SO(7)$:
\be
dim\;[0,1,0]_{B_3}  = 21,
\ee
which agrees with the expected global symmetry. Next, perform the mapping:
 \begin{gather}
  z_4 \rightarrow (z_1 {z_2}^2 z_3^2)^{-1} = x_2^{-1}, \label{eq:E4} \\
   z_1 \rightarrow x_1^2 x_2^{-1}, \; z_2 \rightarrow x_2^2 x_1^{-1} x_3^{-2}, \; z_3 \rightarrow x_3^2 x_2^{-1} ,
 \end{gather}
 such that $z_4$ is eliminated by the gauge fixing condition \eref{eq:E4}. Typically, since $z_4$ is declared to be the null node, it maps to the inverse of the adjoint weight fugacity. The remaining fugacities are mapped to the fundamental weight fugacities of $B_3$ using the Cartan matrix. After the mapping, the refined HS is obtained. For brevity, we only show the $t^2$ coefficient of the expansion of the refined HS:
\begin{gather}
3 + \frac{1}{x_1} + x_1 +  \frac{1}{x_2}+ \frac{x_1}{x_2}+ \frac{x_1^2}{x_2}+ x_2 + \frac{x_2}{x_1^2} +\frac{x_2}{x_1} + 
\frac{x_1}{x_3^2} + \frac{x_2}{x_3^2} +\\
 \frac{x_2}{x_1 x_3^2}+\frac{x_1 x_2}{x_3^2} + \frac{x_2^2}{x_1 x_3^2} +\frac{x_3^2}{x_1}  + \frac{x_1 x_3^2}{x_2^2} + \frac{x_3^2}{x_2}+ \frac{x_3^2}{x_1 x_2}+ \frac{x_1 x_3^2}{x_2}
\end{gather}
which coincides with the character of the adjoint representation of $SO(7)$. This confirms that the global symmetry is $B_3$ and allows us to write the refined PL in the form:
\be \label{eq:plw}
PL=  [0,1,0]_{21}t^2 - ([0,0,0]_1+[0,0,2]_{35}) t^4 +
([1,1,0]_{105} + [0,0,2]_{35})t^6 + O(t^8),
\ee
where $[d_1,d_2,d_3]$ are the Dynkin labels of $B_3$. Recall, that the subscripts denote the dimensions of the corresponding representations. As an algebraic variety, the Coulomb branch is a closure of next to minimal nilpotent orbit of $\mathfrak{so(7)}$ algebra\footnote{This space is defined as a space of $7\times 7$ matrices $M$, satisfying: $M=-M^T,\;Tr(M^2)=0,\;rank(M) \leq2$. In the previous literature, this space is defined with the extra condition $M^3=0$ but Equation \eref{eq:plw} shows that this nilpotency condition is already implied by the rank and the trace conditions.}:
\be \label{eq:MS1}
\mathcal{C}_{[2,1^2]} = \overline{n.min\mathcal{O}_{B_3}}.
\ee
The representation content of the chiral ring is summarized by the HWG in Equation \eref{eq:hwg2}
\begin{equation} \label{eq:hwg2}
HWG=PE[\mu_2 t^2 +\mu_1^2 t^4],
\end{equation}
where $\mu_i,\; i=1,2,3$ are the fugacities for the highest weights of $B_3$. The computation of Equation \eref{eq:hwg2}, which is done starting from the quiver in Figure \eref{fig:D4-2}, provides an independent test that the Coulomb brach moduli space is given by Equation \eref{eq:MS1} since it is consistent with results of Table 10 in \cite{HK16}. The refined analysis together with the fact that the algebraic variety is multiplicity-free determines the Coulomb branch uniquely. The $\mathbb{Z}_2$ quotient between HWG \eref{eq:hwg1} and \eref{eq:hwg2} maps the adjoint rep of $D_4$ into the adjoint and a vector rep of $B_3$. Whereas the adjoint is invariant under this action, the vector transform non-trivially with a minus sign, and hence comes in form of the natural invariant $\mu_1^2 t^4$. Overall, the decomposition of $SO(8)$ into $SO(7)$ can be written as:
\be \label{eq:decomp1}
\mu_2 t^2 \rightarrow \mu_2t^2 + \mu_1^2 t^4
\ee
which is used in the analysis of the next case and the HWG derivation in Appendix \ref{B}.
\subsubsection{Gauging $H_{\lambda} = \mathbb{Z}_2 \times \mathbb{Z}_2$}

\begin{figure}[h!]
\center{
\begin{tikzpicture}[scale=0.70]
\draw (0,0) circle (0.4cm);
\draw (0,-0.8) node {\footnotesize{$2$}};
\draw (0,0) node {\footnotesize{$z_2$}};
\draw (-0.4,0) -- (-0.8,0);
\draw (-1.2,0) circle (0.4cm);
\draw (-1.2,-0.8) node {\footnotesize{$2$}};
\draw (-1.2,0) node {\footnotesize{$z_1$}};
\draw (0.4,0) -- (0.8,0);
\draw (1.2,0) circle (0.4cm);
\draw (1.2,-0.8) node {\footnotesize{$2$}};
\draw (1.2,0) node {\footnotesize{$z_3$}};
 \draw[-,blue] (-0.9,0.3) arc (-66:247:0.7);
 \draw (-1.2,2) node {\footnotesize{$Adj$}};
 \draw[-,blue] (1.5,0.3) arc (-66:247:0.7);
 \draw (1.2,2) node {\footnotesize{$Adj$}};
\end{tikzpicture}
}
\caption{\label{fig:D4-3} $\mathcal{P}_{[2^2]}(4)$ Quiver with $A_3 \cong D_3$ global symmetry, $b=0$, $dim\; \mathcal{M}_C ^\mathbb{H} = 5$.}
\end{figure}
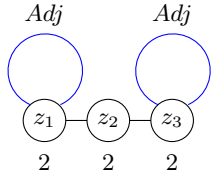

Let us now turn to the construction of the $\mathcal{P}_{[2^2]}(4)$ theory, which is obtained by gauging the subgroup $\mathbb{Z}_2 \times \mathbb{Z}_2 \subset S_4$ of the original permutation symmetry of $\mathcal{P}_{[1^4]}(4)$. According to Conjecture \eref{Main Conjecture}, the desired quiver takes the form depicted, alongside with the assignment of the simple root fugacities, in Figure \eref{fig:D4-3}. The quiver is fully balanced and contains a bouquet of two adjoint $2$ nodes that stems from the central $2$ node.
The anticipated global symmetry on the Coulomb branch is $SU(4) \cong SO(6)$ since the balanced nodes form a Dynkin diagram of $A_3 \cong D_3$. Moreover, $SO(6)$ is also the subgroup which commutes with $\mathbb{Z}_2 \times \mathbb{Z}_2$ inside $SO(8)$. Compute the HS with the simple root fugacities, and set all the fugacities $z_i,\;i=1,2,3$ to unity to find the unrefined HS in Equation \eref{eq:hs1}.
\begin{equation} \label{eq:hs1}
HS_{[2^2]}(t)=\frac{1+10t^2+55t^4+150t^6+288t^8+336t^{10}+288t^{12}+ 
150t^{14}+55t^{16}+10t^{18}+t^{20}}{(1-t^2)^{10}(1+t^2)^5}
\end{equation}
The expansion of the unrefined HS yields
\be
HS(t)=1 + 15 t^2 + 125 t^4 + 685 t^6 + 2898 t^8 + O(t^{10}).
\ee
By taking PL of Equation \eref{eq:hs1} one obtains Equation \eref{eq:uPLx3}.
\be \label{eq:uPLx3}
PL=15 t^2 + 5 t^4 - 70 t^6 + 273 t^8 + O(t^{10}).
\ee
The $t^2$ coefficient in the last equation agrees with the expected global symmetry since the dimension of the adjoint representation of $SU(4)$ is
\be
dim\;[1,0,1]_{A_3} =15.
\ee
In order to perform the refined analysis of the Coulomb branch in terms of the $A_3$ symmetry all three simple root fugacities are needed. This is a complication, however, since eliminating one of the three simple root fugacities by the gauge fixing condition will leave us with just two fugacities to work with. This indicates a presence of certain embedding of a lower rank symmetry inside the $A_3$ global symmetry. 
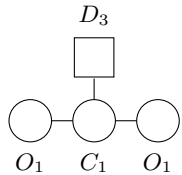
\begin{figure}[h!]
\center{
\begin{tikzpicture}[scale=0.70]
\draw (0,0) circle (0.4cm);
\draw (0,-0.8) node {\footnotesize{$C_1$}};
\draw (-0.4,0) -- (-0.8,0);
\draw (-1.2,0) circle (0.4cm);
\draw (-1.2,-0.8) node {\footnotesize{$O_1$}};
\draw (0.4,0) -- (0.8,0);
\draw (1.2,0) circle (0.4cm);
\draw (1.2,-0.8) node {\footnotesize{$O_1$}};
\draw (0,0.4) -- (0,0.8);
\node at (0,1.2) [minimum size=0.52cm,rectangle,draw] {};
\draw (0,2) node {\footnotesize{$D_3$}};
\end{tikzpicture}
}
\caption{\label{fig:Dual1} Higgs branch quiver with $D_3 \equiv A_3$ global symmetry. Note that $O_1 \equiv \mathbb{Z}_2$ and $C_1$ denotes $Sp(2)$, $dim\; \mathcal{M}_H ^\mathbb{H} = 5$.}
\end{figure}
 In order to do the analysis in terms of the $A_3$, observe that the unrefined HS computed in Equation \eref{eq:hs1} is also the unrefined HS for a Higgs branch quiver in Figure \eref{fig:Dual1}, where gauge (round) and flavor (square) groups are denoted explicitly\footnote{The first computation of the Higgs branch quiver was done by Rudolph Kalveks.}. Note that the $O_1$ nodes (or equivalently $\mathbb{Z}_2$) precisely realize the two $\mathbb{Z}_2$ actions on the Higgs branch of a $Sp(1)$ gauge theory with a $D_4$ flavor group and by gauging the two factors of $\mathbb{Z}_2$ in the global symmetry, one recovers the quiver depicted in Figure \eref{fig:Dual1}, where the remaining global symmetry is $SO(6) \cong SU(4)$. Given the above motivations, let us bypass the problem constituted by the missing fugacities in Figure \eref{fig:D4-3} and use a computation of the Higgs branch of the quiver in Figure \eref{fig:Dual1} instead. After appropriate fugacity maps one can show that the refined Hilbert series are equal to each other. Let us start with the quiver in Figure \eref{fig:D4-3}. After the computation of the HS using simple root fugacities $z_i, i=1,2,3$ impose the gauge fixing condition:
 \begin{gather}
z_3 \rightarrow (z_1^2 z_2^2)^{-1/2}
 \end{gather}
 eliminating the $z_3$ fugacity. Recall, that the gauge fixing follows from constraint \eref{eq:constraint1}. The obtained HS now only contains $z_1$ and $z_2$ fugacities. The $t^2$ coefficient of the refined HS takes the form
\be \label{eq:observ1}
3 + \frac{2}{z_1} + 2 z_1 + \frac{1}{z_2} + \frac{1}{z_1^2 z_2} + \frac{2}{z_1 z_2} + z_2 + 2 z_1 z_2 + 
 z_1^2 z_2.
 \ee
 Written in terms of the simple roots, this is precisely the character of the adjoint representation of $A_3$ under the identification $z_3 \rightarrow z_1$!\\

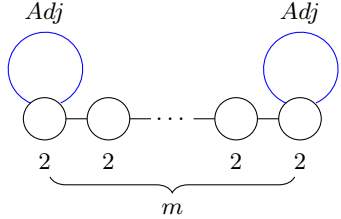
\begin{figure}[h!]
\center{
\begin{tikzpicture}[scale=0.70]
\draw (0,0) node {\footnotesize{$\dots$}};
\draw (-0.4,0) -- (-0.8,0);
\draw (-1.6,0) -- (-2,0);
\draw (1.6,0) -- (2,0);
\draw (-1.2,0) circle (0.4cm);
\draw (-1.2,-0.8) node {\footnotesize{$2$}};
\draw (0.4,0) -- (0.8,0);
\draw (1.2,0) circle (0.4cm);
\draw (1.2,-0.8) node {\footnotesize{$2$}};
\draw (2.4,0) circle (0.4cm);
\draw (2.4,-0.8) node {\footnotesize{$2$}};
\draw (-2.4,0) circle (0.4cm);
\draw (-2.4,-0.8) node {\footnotesize{$2$}};
 \draw[-,blue] (-2.1,0.3) arc (-66:247:0.7);
 \draw (-2.4,2) node {\footnotesize{$Adj$}};
 \draw[-,blue] (2.7,0.3) arc (-66:247:0.7);
 \draw (2.4,2) node {\footnotesize{$Adj$}};
\draw [decorate,decoration={brace,amplitude=6pt}] (2.3,-1.1) to (-2.3,-1.1);
\draw (0,-1.7) node {\footnotesize{$m$}};
\end{tikzpicture}
}
\caption{\label{fig:Cmirror} Coulomb branch quiver with $A_m$ global symmetry, $dim\; \mathcal{M}_C ^\mathbb{H} = 2m-1$.}
\end{figure}

On the other hand, the HWG for the Higgs branch quiver in Figure \eref{fig:Dual1} is given by Equation \eref{eq:hwg-22}
\be \label{eq:hwg-22}
HWG=PE[\mu_1 \mu_3 t^2+({2\mu_2}^2  +1)t^4 + {\mu_2}^2 t^6 - {\mu_2}^4 t^{12}],
\ee
where $\mu_i,\; i=1,2,3$ are the highest weight fugacities of $A_3$. The derivation of HWG \eref{eq:hwg-22} is included in Appendix \ref{B}. One can turn this HWG into the refined Hilbert series which is expressed using the fundamental weight fugacities $x_i, i=1,2,3$. Further, lets use the inverse of the Cartan matrix to map the $x_i$ fugacities in the refined HS to the simple root fugacities $z_i, i=1,2,3$. The desired fugacitiy map takes the form:
 \begin{gather}
 x_1 \rightarrow (z_1^3 z_2^2 z_3)^{\frac{1}{4}}, \; x_2 \rightarrow  (z_1 z_2^2 z_3)^{\frac{1}{2}} ,\; x_3 \rightarrow (z_1 z_2^2 z_3^3 )^{\frac{1}{4}} .
 \end{gather}
 At this stage, the refined HS for the Higgs branch quiver is expressed using all three simple root fugacities. As a final step make the same identification used to recover the correct character in front of $t^2$ coefficient in Equation \eref{eq:observ1}. Recall, the form of the identification:
 \be
 z_3 \rightarrow z_1.
 \ee
 Finally, the two Hilbert series, obtained by working from both sides of the duality and using the fugacity maps prescribed above, are equal! This verifies that the global symmetry of the $\mathcal{P}_{[2^2]}$ theory in Figure \eref{fig:D4-3} is $A_3$ and the chiral ring is described by the HWG in Equation \eref{eq:hwg-22}. By the sequel, the PL of the refined HS can be written in the form:
\be
\begin{split}
PL&= [1,0,1]_{15}t^2 + ([0,2,0]_{20} - [1,0,1]_{15}) t^4 - ([0,2,0]_{20}-[2,1,0]_{45}-[0,1,2]_{45}) t^6 \\
&+  ([2,1,0]_{45}+2[1,0,1]_{15}+[1,2,1]_{175}+[0,1,2]_{45}-[0,2,0]_{20}-2) t^8 + O(t^{10}),
\end{split}
\ee
where $[d_1,d_2,d_3]$ are the Dynkin labels of the $A_3$ representations. Recall, that the subscripts denote the dimensions of the corresponding representations. Consider a quiver consisting of a chain of $m$ rank $2$ nodes such that the two boundary nodes are adjoint. The quiver is depicted in Figure \eref{fig:Cmirror}. A generalization of the last derivation, supported by computational evidence up to $m=4$, implies that the following conjecture holds: \\

\begin{figure}[h!]
\center{
\begin{tikzpicture}[scale=0.70]
\draw (0,0) circle (0.4cm);
\draw (0,-0.8) node {\footnotesize{$C_1$}};
\draw (-0.4,0) -- (-0.8,0);
\draw (-1.2,0) circle (0.4cm);
\draw (-1.2,-0.8) node {\footnotesize{$O_1$}};
\draw (0.4,0) -- (0.8,0);
\draw (1.2,0) circle (0.4cm);
\draw (1.2,-0.8) node {\footnotesize{$O_1$}};
\draw (0,0.4) -- (0,0.8);
\node at (0,1.2) [minimum size=0.52cm,rectangle,draw] {};
\draw (0,2) node {\footnotesize{$D_m$}};
\end{tikzpicture}
}
\caption{\label{fig:Hmirror} Higgs branch quiver with $D_m$ global symmetry, $dim\; \mathcal{M}_H ^\mathbb{H} = 2m-1$.}
\end{figure}
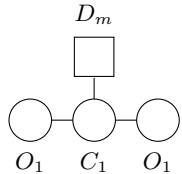

\emph{The Coulomb branch of quiver in Figure \eref{fig:Cmirror} is equal to the Higgs branch of the quiver in Figure \eref{fig:Hmirror}.}


One can consider a quiver for an $Sp(1)$ gauge theory with $m$ flavors and realize the two $\mathbb{Z}_2$ actions. The same can be done for the Coulomb branch quiver in the form of the affine $D_{m+2}$ Dynkin diagram such that the fork $U(1)$ nodes on both ends are substituted by adjoint $2$ nodes (due to the two $\mathbb{Z}_2$ actions). The obtained Higgs and Coulomb branch quivers are precisely those in Figures \eref{fig:Cmirror} and \eref{fig:Hmirror}.

\subsubsection{Gauging $H_{\lambda} = S_3$}

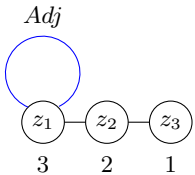
\begin{figure}[h!]
\center{
\begin{tikzpicture}[scale=0.70]
\draw (0,0) circle (0.4cm);
\draw (0,-0.8) node {\footnotesize{$2$}};
\draw (0,0) node {\footnotesize{$z_2$}};
\draw (-0.4,0) -- (-0.8,0);
\draw (-1.2,0) circle (0.4cm);
\draw (-1.2,-0.8) node {\footnotesize{$3$}};
\draw (-1.2,0) node {\footnotesize{$z_1$}};
\draw (0.4,0) -- (0.8,0);
\draw (1.2,0) circle (0.4cm);
\draw (1.2,-0.8) node {\footnotesize{$1$}};
\draw (1.2,0) node {\footnotesize{$z_3$}};
 \draw[-,blue] (-0.9,0.3) arc (-64:247:0.7);
 \draw (-1.2,2) node {\footnotesize{$Adj$}};
\end{tikzpicture}
}
\caption{\label{fig:D4-4} $\mathcal{P}_{[3,1]}(4)$ Quiver with $G_2$ global symmetry, $dim\; \mathcal{M}_C ^\mathbb{H} = 5$.}
\end{figure}

The next theory is obtained by gauging an $S_3$ subgroup of the $S_4$ discrete global symmetry of the parent $\mathcal{P}_{[1^4]}(4)$ quiver. Conjecture \eref{Main Conjecture} implies that the result of such discrete gauging produces the $\mathcal{P}_{[3,1]}(4)$ quiver, depicted in Figure \eref{fig:D4-4}. The assignment of the simple root fugacities is also shown in Figure \eref{fig:D4-4}. The anticipation of the global symmetry follows from the comparison of this quiver to an affine $G_2$ Dynkin diagram in a similar fashion as in the case of the $\mathcal{P}_{[2,1^2]}(4)$ quiver. Moreover, $G_2$ is also the a subgroup that commutes with $S_3$ inside $SO(8)$.
We proceed by computing the HS and unrefining by setting all simple root fugacities, $z_i, i= 1,2,3$, to $1$. The unrefined HS is given by Equation \eref{eq:hs2}.
\begin{equation} \label{eq:hs2}
HS_{[3,1]}(t)=\frac{(1 + t^2) (1 + 3 t^2 + 6 t^4 + 3 t^6 + t^8)}{(1-t^2)^{10}}
\end{equation}
The expansion of the unrefined HS reads
\be
HS_{[3,1]}(t)=1 + 14 t^2 + 104 t^4 + 539 t^6 + 2184 t^8 +O(t^{10}).
\ee
Note, that \eref{eq:hs2} agrees with the result of the HS for the sub-regular nilpotent orbit of $G_2$ in Table 3 in \cite{HKexceptionals17}. The unrefined PL takes the form
\be
PL=14 t^2 - t^4 - 7 t^6 + 7 t^8 + O(t^9).
\ee
The $t^2$ coefficient of the PL is the dimension of the adjoint representation of $G_2$:
\be
dim\; [1,0]_{G_2} = 14 .
\ee
 In fact, the global symmetry for the quiver in Figure \eref{fig:D4-4} is argued to be $G_2$ in \cite{GR12}. In order to confirm this expectation on a level of the refined HS, the following mappings need to be employed.

\paragraph{Mapping of $\mathcal{P}_{[3,1]}(4)$ simple root fugacities to the highest weight fugacities of $G_2$:}

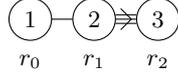
\begin{figure}[h!]
\center{
\begin{tikzpicture}[scale=0.70]
\draw (0,0) circle (0.4cm);
\draw (0,-0.8) node {\footnotesize{$r_1$}};
\draw (0,0) node {\footnotesize{$2$}};
\draw (-0.4,0) -- (-0.8,0);
\draw (-1.2,0) circle (0.4cm);
\draw (-1.2,-0.8) node {\footnotesize{$r_0$}};
\draw (-1.2,0) node {\footnotesize{$1$}};
\draw (0.4,0) -- (0.8,0);
\draw (0.4,0.08) -- (0.8,0.08);
\draw (0.4,-0.08) -- (0.8,-0.08);
\draw (0.5,0.2) -- (0.7,0);
\draw (0.5,-0.2) -- (0.7,0);
\draw (1.2,0) circle (0.4cm);
\draw (1.2,-0.8) node {\footnotesize{$r_2$}};
\draw (1.2,0) node {\footnotesize{$3$}};
\end{tikzpicture}
}
\caption{\label{fig:g2aff} Affine Dynkin diagram of $G_2$ with Coxeter labels and simple root fugacities.}
\end{figure}

The procedure involves three steps. First step is to impose the usual gauge condition which eliminates the fugacity of the adjoint $3$ node\footnote{One could equally eliminate the rank $1$ node and adjust for such change in the next mappings.}. The map is given by Equation \eref{eq:fm-232}
\be
  z_1 \rightarrow (z_2^2 z_3)^{- \frac{1}{3}}, \label{eq:fm-232}
  \ee
which leaves us with fugacities $z_2$ and $z_3$. Second step is to map these fugacities to the simple root fugacities of $G_2$ (i.e. one needs to find a Dynkin map from $A_2$ to $G_2$). For this purpose, consider the affine Dynkin diagram of $G_2$, depicted in Figure \eref{fig:g2aff}. The Coxeter labels are indicated inside the nodes.
In the Dynkin map, the $r_0$ fugacity of the affine node will play no role, hence we can express $r_0$ in terms of the other two:
\be
r_0 = (r_1^2 r_2^3)^{-1}
\ee
Note that all fugacities are weighted by their Coxeter labels. Further, by comparing Figures \eref{fig:D4-4} and \eref{fig:g2aff}, one sees that, in the Dynkin map, $z_2$ should map to $r_1$. Moreover, the $z_3$ fugacity maps to $r_0$ fugacity of the affine node as these are the corresponding rank $1$ nodes. Hence, the desired Dynkin map is:
 \begin{gather}
 z_2 \rightarrow r_1 \\
 z_3 \rightarrow r_0 = (r_1^2 r_2^3),
 \end{gather}
which concludes the second step. The last step is to map the $G_2$ simple root fugacities $r_1, r_2$ to the coordinates on the weight space of $G_2$. Employing the Cartan matrix of $G_2$ one finds that Equations \eref{eq:g2r2c1} and \eref{eq:g2r2c2} provide the desired map.
 \begin{gather}
 r_1 \rightarrow y_1^2 y_3^{-3} \label{eq:g2r2c1} \\
 r_2 \rightarrow y_2^2 y_1^{-1}. \label{eq:g2r2c2}
 \end{gather}
After these mappings, the $t^2$ coefficient of the refined HS is computed as
\be
2 + \frac{1}{y_1} + y_1 + \frac{y_1}{y_2^3}+\frac{y_1^2}{y_2^3} +\frac{y_1}{y_2^2}  +\frac{1}{y_2}+\frac{y_1}{y_2} + y_2 +
\frac{y_2}{y_1} +\frac{y_2^2}{y_1} +\frac{y_2^3}{y_1^2} +\frac{y_2^3}{y_1},
\ee
which is precisely the character of the $14$ dimensional adjoint representation of $G_2$. Thus, the expectation of global symmetry is verified and in agreement with arguments in \cite{GR12}. Finally, one can write the refined PL in the form:
 \begin{equation} \label{eq:pl}
PL= [1,0]_{14}t^2 - [0,0]_1 t^4 - [0,1]_7 t^6 +[0,1]_7 t^8+ O(t^{10}).
\end{equation}
There are two relations at: $t^4$ transforming as a singlet, and at $t^6$ transforming under the $7$ dimensional $[0,1]$ representation of $G_2$, respectively.  
These relations can be summarized by the algebraic variety made out of 14 complex numbers $M^a$, in the adjoint representation of $G_2$, which satisfy the relations
$$M^a M^a = 0,$$ and $$M M M |_{[0,1]} = 0.$$
The Coulomb branch of the $\mathcal{P}_{[3,1]}(4)$ theory is the $10$ dimensional sub-regular nilpotent orbit of $G_2$ \cite{HKexceptionals17}:
\be
\mathcal{C}_{[3,1]}= \overline{sub.reg.\mathcal{O}_{G_2}}. 
\ee
 The formula for the HWG is given by Equation \eref{eq:hwg-23}, which is Equation (3.37) in \cite{Sic14}, where the authors used a different convention for the factor multiplying the conformal dimension in the monopole formula (i.e. all $t$ powers are half of those herein).
\be \label{eq:hwg-23}
HWG=PE[\mu_2 t^2 +{\mu_1}^2 t^4 +{\mu_1}^3 t^6 +{\mu_2}^2 t^8 +{\mu_1}^3 \mu_2 t^{10}-{\mu_1}^6 {\mu_2}^2 t^{20}]
\ee
Let us refer to the quiver in Figure \eref{fig:D4-4} as the \textit{$G_2$-tail} for the following reason. Consider a construction defined by two steps:
\begin{itemize}
\item Consider any quiver $\sf Q$, with $G_0$ global symmetry and attach the $G_2$-tail (which has a $G_2$ global symmetry) to this quiver via an adjoint node\footnote{In case of $A$-series, one needs to attach the $G_2$-tail via both of the adjoint nodes.}
\item Multiply all the ranks of $\sf Q$ by $3$
\end{itemize}
Then, the theory constructed by this procedure has a global symmetry $G_{global}$ that satisfies Equation (\ref{eq:Glob}).
\be \label{eq:Glob}
G_{global}=G_0 \times G_2
\ee
For detailed examples of this construction, see (5.24) and (5.25) and the consequent discussion in \cite{MOTZ17}.

\subsubsection{Gauging $H_{\lambda} = S_4$}

\begin{figure}[h!]
\center{
\begin{tikzpicture}[scale=0.70]
\draw (0,0) circle (0.4cm);
\draw (0,-0.8) node {\footnotesize{$2$}};
\draw (0,0) node {\footnotesize{$z_2$}};
\draw (0,0.4) -- (0,0.8);
\draw (0,1.2) circle (0.4cm);
\draw (0,2) node {\footnotesize{$4$}};
\draw (0,1.2) node {\footnotesize{$z_1$}};
 \draw[-,blue] (0.335,1.455) arc (-64:247:0.8);
 \draw (0,3.27) node {\footnotesize{$Adj$}};
\end{tikzpicture}
}
\caption{\label{fig:D4-5} $\mathcal{P}_{[4]}(4)$ Quiver with $A_2$ global symmetry, $dim\; \mathcal{M}_C ^\mathbb{H} = 5$.}
\end{figure}
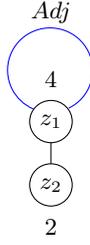

In order to construct the last theory, gauge the entire discrete $S_4$ symmetry of the parent $\mathcal{P}_{[1^4]}(4)$ quiver. According to Conjecture \eref{Main Conjecture}, the desired $\mathcal{P}_{[4]}(4)$ quiver takes the form depicted in Figure \eref{fig:D4-5}. The assignment of the simple root fugacities is also shown in Figure \eref{fig:D4-5}.
The balanced part of this quiver forms the $A_2$ Dynkin diagram therefore the expected global symmetry is $SU(3)$. It is also the subgroup that commutes with $S_4$ inside $SO(8)$. Moreover, as in the case of the $\mathcal{P}_{[2^2]}$ quiver, the number of fugacities after the gauge fixing is smaller than the rank of the expected global symmetry. Thus, one expects to find a certain embedding: $SU(2) \hookleftarrow SU(3)$. In fact,  the set of all embedding of $\mathfrak{su}(2)$ inside a $\mathfrak{su}(n)$ algebra is in one-to-one correspondence with the set of all nilpotent orbits of $\mathfrak{su}(n)$ and there is a bijection between nilpotent orbits and the partitions of $\mathcal{P}(n)$. As a first step, the HS is computed using the simple root fugacities, and then unrefined by setting all fugacities to unity. The unrefined HS is given by Equation \eref{eq:hs5}.
\begin{equation} \label{eq:hs5}
HS_{[4]}(t)=\frac{1+3t^2+13t^4+25t^6+46t^8+48t^{10}+46t^{12}+25t^{14}+
 13t^{16}+3t^{18}+t^{20}}{(1-t^2)^{5}(1 - t^4)^5}.
\end{equation}
The expansion of the unrefined Hilbert series reads
\be
HS_{[4]}(t)=1 + 8 t^2 + 48 t^4 + 210 t^6 + 771 t^8 + O(t^{10}).
\ee
 The expression of the unrefined PL takes the form
 \be
 PL= 8 t^2 + 12 t^4 - 6 t^6 - 21 t^8 + O(t^{10}).
 \ee
The $t^2$ coefficient agrees with the dimension of the adjoint representation of $SU(3)$:
 \be
 dim\; [1,1]_{A_2} = 8
 \ee
 Lets proceed by mapping the simple root fugacities $z_i, i =1,2$ according to \eref{eq:fm-241} and \eref{eq:fm-242}
 \begin{gather}
 z_2 \rightarrow ({z_1}^4)^{-1/2} \label{eq:fm-241}  \\
 z_1 \rightarrow x^2 \label{eq:fm-242} ,
 \end{gather}
 where $z_2$ fugacity of the rank $2$ node is eliminated by constraint \eref{eq:constraint1}, and $x$ is the fugacity for the fundamental weight of $SU(2)$. Using this mapping, the expansion of the refined PL takes the form:
 \be
 \begin{split} \label{eq:embpl1}
PL&= \left( 2 + \frac{1}{x^4} + \frac{2}{x^2} + 2 x^2 + x^4\right)t^2 + \left( 4+\frac{2}{x^4} +\frac{2}{x^2} + 2x^2 + 2x^4 \right)t^4 \\
&- \left(2+ \frac{2}{x^2} + 2x^2 \right)t^6 + \left(7+ \frac{3}{x^4}+\frac{4}{x^2}+4x^2 +3x^4\right)t^8 + O(t^9).
 \end{split}
 \ee
 The last expression can be written as:
 \be
 PL= ([4]_5+[2]_3)t^2 + (2[4]_5+2[0]_1)t^4 - (2[2]_3) t^6 -(3[4]_5+[2]_3+3[0]_1)t^8 + O(t^9)
 \ee
where $[a]$ is used to denote the Dynkin labels of $SU(2)$ representations. One can list representations at each order of $t$ as follows:
 \begin{itemize}
  \item $t^2$: generators transforming under $[4]+[2]$
  \item $t^4$: generators transforming under $2[4]+2[0]$
  \item $t^6$: relations transforming under $[2]$
  \item $t^8$: relations transforming under $3[4] +[2]+ 3\times[0]$
 \end{itemize}
The obtained embedding of $SU(2)$ inside $SU(3)$ corresponds to the homomorphism embedding characterizing the maximal nilpotent orbit of $SU(3)$ (f.i. see the last row of second Table in Appendix B.1 in \cite{HK16}):
\be
[4]_5 \oplus [2]_3 \hookleftarrow [1,1]_8
\ee
This provides a verification of the expected global symmetry since the $\mathfrak{su}(2)$ embedding in case of the maximal nilpotent orbit of $\mathfrak{su}(3)$ is characterized by a map where the two fugacities of the $A_2$, $(y_1, y_2)$ map to $(x^2,1)$, where $x$ is the $SU(2)$ fugacity, and so in turn, the character of the adjoint representation of $A_3$ $[1,1]$ becomes the character of $[4]\oplus [2]$ of $A_1$, which is the $t^2$ coefficient in Equation \eref{eq:embpl1}. In terms of the $A_2$ Dynkin labels, the refined PL can be written in the form:
\be \label{eq:pl-24}
PL= [1,1]_8 t^2 + ([2,0]_6 +[0,2]_6 ) t^4 - ( [1,0]_3 +[0,1]_3 ) t^6 -([1,1]_{15} + [2,0]_6 + [0,2]_6 +1)t^8 +O(t^9).
\ee
The eight generators of the global symmetry transform under the adjoint representation of $A_2$. At order $t^4$ there are generators transforming under the $[2,0]$ and the conjugate $[0,2]$ representation. The relation at order $t^6$ transforms under the fundamental and anti-fundamental representations denoted by $[1,0]$ and $[0,1]$, respectively. Finally, the relations at $t^8$ transform under $[1,1]+[2,0]+[0,2]+[0,0]$.

\subsubsection{Comparison of the Coulomb branch volumes}

Consider the unrefined HS computed in this subsection for all five theories. Recall that these are: \eref{eq:hs21}, \eref{eq:hs22}, \eref{eq:hs1}, \eref{eq:hs2} and \eref{eq:hs5}. For each pair of theories, expand the unrefined Hilbert series according to Equation \eref{eq:expansionHS} and plug into Equation \eref{eq:01}. The computed ratios of the Coulomb branch volumes of $k=n=2$ theories are summarized in Table \eref{tab:Ratios1}. 
\begin{table}
\begin{center}
\begin{tabular}{ |p{1.6cm}||p{1.5cm}|p{1.5cm}|p{1.5cm}|p{1.5cm}|p{1.5cm}| }
 \hline
 \multicolumn{6}{|c|}{Ratios of Coulomb branch volumes for $n=2$, $k=2$ theories} \\
 \hline
 Partition & $[1^4]$ &$[2,1^2]$&$[2^2]$&$[3,1]$&$[4]$\\
 \hline \hline
 $[1^4]$&1&2&4&6&24\\
 $[2,1^2]$&&1&2&3&12\\
$[2^2]$&&&1&3/2&6\\
 $[3,1]$&&&&1&4\\
 $[4]$&&&&&1\\
 \hline
\end{tabular}
\end{center}
\caption{\label{tab:Ratios1} Ratios of Coulomb branch volumes for $n=2$, $k=2$ theories.}
\end{table}
The parent Coulomb branch of $\mathcal{P}_{[1^4]}(4)$ is $ \overline{min\mathcal{O}_{D_4}}$. All the evidence for Conjecture \eref{Main Conjecture} suggests that for the daughter Coulomb branches, there holds:
\begin{equation}
\mathcal{C} = \overline{min \mathcal{O}_{D_4}}/ \Gamma
\end{equation}
where $\Gamma \subseteq S_4$ is a discrete group. In particular:
\be
\Gamma = \left\{\begin{array}{l}S_2 \equiv \mathbb{Z}_2 \quad for \;[2,1^2]\\ \mathbb{Z}_2 \times \mathbb{Z}_2 \quad for\; [2^2]\\ S_3 \quad \quad \quad \; for \;[3,1]\\ S_4 \quad \quad \quad \; for \;[4]\end{array}\right.
\ee
Note that the obtained relations \eref{eq:NOrelation1} and \eref{eq:NOrelation2}

\begin{gather} \label{eq:NOrelation1}
\overline{min \mathcal{O}_{D_4}}/ \mathbb{Z}_2=\overline{n.min \mathcal{O}_{B_3}} \\
\overline{min \mathcal{O}_{D_4}}/ S_3=\overline{sub.reg \mathcal{O}_{G_2}} \label{eq:NOrelation2} 
\end{gather}
relate quotients of the closure of the minimal nilpotent orbit of $\mathfrak{so(8)}$ algebra to the (closures of the) next to minimal orbit of $\mathfrak{so(7)}$ and sub-regular nilpotent orbit of $\mathfrak{g_2}$, respectively. This is the classic result of Kostant and Brylinski \cite{KB92}.


\section{Second Family: Bouquet quivers with ${A_1}^{n_1}\times D_{n_2 +1}$ global symmetry} \label{2}

The analysis of the first family of bouquet quivers allows a generalization of the results of the HWG to a larger family of quivers. In the first part of this section we derive a general formula of the HWG for the \emph{second family} of quivers with ${A_1}^{n_1} \times D_{n_2 +1}$ global symmetry. In the second part of the section, we perform discrete gauging for a particular member of this family. The main focus in this section is the derivation of the general formula for the HWG and the analysis of the particular quivers is given in less detail. The simple root fugacities, indicated inside the quiver nodes in the figures, are denoted by $z_i$ and $w_i$. The fugacity maps throughout this section show the mappings of the simple root fugacities to the fundamental weight fugacities which are denoted by $x_i$ and $y_i$. Finally, the highest weight fugacities used in the expressions of HWG are denoted by $\mu_i$ and $\nu_i$. 

\subsection{Derivation of HWG}

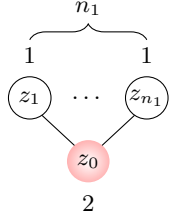
\begin{figure}[h!]
\center{
\begin{tikzpicture}[scale=0.70]
\shade[inner color=white,outer color=red!30] (0,0) circle (0.4cm);
\draw (0,-0.8) node {\footnotesize{$2$}};
\draw (0,0) node {\footnotesize{$z_0$}};
\draw (-0.26,0.3) -- (-0.87,0.87);
\draw (-1.1,1.2) circle (0.4cm);
\draw (-1.1,2) node {\footnotesize{$1$}};
\draw (-1.1,1.2) node {\footnotesize{$z_1$}};
\draw (0.26,0.3) -- (0.87,0.87);
\draw (1.1,1.2) circle (0.4cm);
\draw (1.1,2) node {\footnotesize{$1$}};
\draw (1.1,1.2) node {\footnotesize{$z_{n_1}$}};
\draw (0,1.2) node {\footnotesize{$\dots$}};
\draw [decorate,decoration={brace,amplitude=6pt}] (-1.1,2.3) to (1.1,2.3);
\draw (0,2.9) node {\footnotesize{$n_1$}};
\end{tikzpicture}
}
\caption{\label{fig:DGen1} $\mathcal{P}_{[1^{n_1}]}(n_1)$ Quiver, $b=n_1-4$, $dim \mathcal{M}_C ^\mathbb{H} = n_1 +1$.}
\end{figure}

Consider the theory in Figure \eref{fig:4} and set $k=2$, $n=n_1$. The corresponding quiver is depicted in Figure \eref{fig:DGen1}. The central node is balanced for a special case $n_1 = 4$, which is indicated by the radial color gradient of the node. The theory in Figure \eref{fig:DGen1} has a $SU(2)^{n_1}$ global symmetry which enhances\footnote{See section \ref{1}.} to $SO(8)$ for $k_1=4$.

 For $n_1 \neq 4$ all the bouquet nodes are balanced and the only unbalanced node is the central one, with balance $b=n_1 -4$. Lets consider the $n_1 =5$ case. The balanced sub-quivers form five $A_1$ Dynkin diagrams, therefore the expected global symmetry is $SU(2)^5$. Analogically to Figure \eref{fig:D4-1} ,the simple root fugacities are assigned such that $z_0$ is the simple root fugacity of the unbalanced node and $z_i, i= 1, \dots, 5$ are the simple root fugacities of the bouquet nodes. Computation of the unrefined HS yields 
 \be
 HS(t)=\frac{P_1(t)}{(1-t)^{12} (1+t)^4 (1+t+t^2)^6}
 \ee
 where
 \be
 \begin{split}
 P_1(t)&= 1 - 2 t + 12 t^2 + 4 t^3 + 21 t^4 + 60 t^5 + 54 t^6 + 66 t^7 \\
 &+120 t^8 +\cdots palindrome \cdots + t^{16}.
 \end{split}
 \ee
 The expansion of the unrefined HS is given by Equation \eref{eq:exp1}
 \be \label{eq:exp1}
 HS(t)=1 + 15 t^2 + 32 t^3 + 116 t^4 + 352 t^5 + 863 t^6 + 2112 t^7 + O(t^8).
 \ee
 The PL of the unrefined HS is computed as
 \be
 PL=15 t^2 + 32 t^3 - 4 t^4 - 128 t^5 - 285 t^6 + 320 t^7 + O(t^8).
 \ee
 The $t^2$ coefficient of the last expression is the dimension of the expected $G_{global}$:
 \be
 5 \times dim\; [2]_{A_1} = 15.
 \ee
 Perform the mapping according to \eref{eq:FMG1}, \eref{eq:FMG2} and \eref{eq:FMG3}
\begin{gather}
 z_i\rightarrow x_i^2,\; i=1,2, 3 \label{eq:FMG1} \\ 
 z_4\rightarrow y_1^2,\; z_5 \rightarrow y_2^2 \label{eq:FMG2} \\ 
 z_0\rightarrow (z_1 z_2 z_3 z_4 z_5)^{-\frac{1}{2}} =(x_1 x_2 x_3 y_1 y_2)^{-1} .  \label{eq:FMG3}
 \end{gather}
such that the unbalanced fugacity $z_0$ is eliminated\footnote{Recall, that the elimination follows from \eref{eq:constraint1}.} and ${x_i}$ and ${y_i}$ are the fundamental weight fugacities of $SU(2)$. Note the splitting of fugacities of the bouquet nodes into $x_1, x_2, x_3$ and $y_1, y_2$. The reason for such splitting will shortly become apparent. After the computation of the refined HS one makes use of the HWG to describe the chiral ring of the theory. The HWG takes the form given by Equation \eref{eq:G1} \cite{HM11}:
\be \label{eq:G1}
 \begin{split}
HWG=& PE[ \left(\nu_1 ^2 +\nu_2^2+\nu_3^2 + \mu_1^2+\mu_2^2\right)t^2 + \left(\nu_1\nu_2\nu_3\mu_1\mu_2\right)\left(t^3+t^5\right)  \\
&+ \left(\mu_1^2\mu_2^2\right)t^4 -\left(\mu_1^2\mu_2^2\right)t^4 +t^4 - \left(\nu_1\nu_2\nu_3\mu_1\mu_2\right)^2 t^{10}],
\end{split}
\ee
where $\mu_i,\;i=1,2$ and $\nu_i,\;i=1,2,3$ are the fugacities for the highest weights of $SU(2)$. The $t^2$ terms in Equation \eref{eq:G1} are the usual contributions of the global symmetry for each of the balanced $SU(2)$ nodes. The imbalance of the central node, $b=1$, produces the $t^3$ contribution in Equation \eref{eq:G1}. Furthermore, since the $5$ bouquet nodes are conected to the unbalanced node, the resulting operators transform in the multi-fundamental representation corresponding to all of bouquet nodes, denoted by $\nu_1 \nu_2 \mu_1 \mu_2 \mu_3$. The $t^5$ naturally comes from the tensor product of the adjoint and the multi-fundamental representation. It should also be emphasized that a zero in the form of $(\mu_1^2 \mu_2^2)t^4-(\mu_1^2 \mu_2^2)t^4$ is added to expression \eref{eq:G1} in anticipation of the $\mu_1^2 t^4$ term of \eref{eq:G2}. The $t^4$ singlet term shows up since the Casimir invariant of the five $SU(2)$ are all proportional to each other. Finally, there is a relation at $t^{10}$ transforming under $[2;2;2;2;2]$ (i.e. the adjoint five-representation of $SU(2)^5$). \\

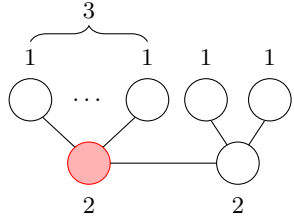
\begin{figure}[h!]
\center{
\begin{tikzpicture}[scale=0.70]
\draw (0,0)[red,fill=red!30] circle (0.4cm);
\draw (0,-0.8) node {\footnotesize{$2$}};
\draw (2.8,0) circle (0.4cm);
\draw (2.8,-0.8) node {\footnotesize{$2$}};
\draw (0.4,0) -- (2.4,0);
\draw (-0.26,0.3) -- (-0.87,0.87);
\draw (-1.1,1.2) circle (0.4cm);
\draw (-1.1,2) node {\footnotesize{$1$}};
\draw (0.26,0.3) -- (0.87,0.87);
\draw (1.1,1.2) circle (0.4cm);
\draw (1.1,2) node {\footnotesize{$1$}};
\draw (0,1.2) node {\footnotesize{$\dots$}};
\draw [decorate,decoration={brace,amplitude=6pt}] (-1.1,2.3) to (1.1,2.3);
\draw (0,2.9) node {\footnotesize{$3$}};
\draw (2.6,0.34) -- (2.3,0.8);
\draw (2.2,1.2) circle (0.4cm);
\draw (3,0.34) -- (3.3,0.8);
\draw (3.4,1.2) circle (0.4cm);
\draw (2.2,2) node {\footnotesize{$1$}};
\draw (3.4,2) node {\footnotesize{$1$}};
\end{tikzpicture}
}
\caption{\label{fig:DGen2} $\{\mathcal{P}_{[1^3]}(3)$, $\mathcal{P}_{[1^2]}(2)\}$ Quiver with $SU(2)^{3}\times D_3$, global symmetry, $b=1$, $dim\; \mathcal{M}_C ^\mathbb{H} =8$.}
\end{figure}

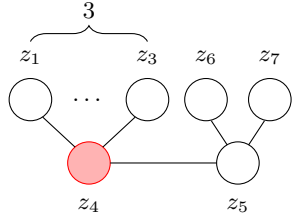
\begin{figure}[h!]
\center{
\begin{tikzpicture}[scale=0.70]
\draw (0,0)[red,fill=red!30] circle (0.4cm);
\draw (0,-0.8) node {\footnotesize{$z_4$}};
\draw (2.8,0) circle (0.4cm);
\draw (2.8,-0.8) node {\footnotesize{$z_5$}};
\draw (0.4,0) -- (2.4,0);
\draw (-0.26,0.3) -- (-0.87,0.87);
\draw (-1.1,1.2) circle (0.4cm);
\draw (-1.1,2) node {\footnotesize{$z_1$}};
\draw (0.26,0.3) -- (0.87,0.87);
\draw (1.1,1.2) circle (0.4cm);
\draw (1.1,2) node {\footnotesize{$z_3$}};
\draw (0,1.2) node {\footnotesize{$\dots$}};
\draw [decorate,decoration={brace,amplitude=6pt}] (-1.1,2.3) to (1.1,2.3);
\draw (0,2.9) node {\footnotesize{$3$}};
\draw (2.6,0.34) -- (2.3,0.8);
\draw (2.2,1.2) circle (0.4cm);
\draw (3,0.34) -- (3.3,0.8);
\draw (3.4,1.2) circle (0.4cm);
\draw (2.2,2) node {\footnotesize{$z_6$}};
\draw (3.4,2) node {\footnotesize{$z_7$}};
\end{tikzpicture}
}
\caption{\label{fig:DGen2fugs} $\{\mathcal{P}_{[1^3]}(3)$, $\mathcal{P}_{[1^2]}(2)\}$ Quiver with simple root fugacities.}
\end{figure}

\begin{figure}[h!]
\center{
\begin{tikzpicture}[scale=0.70]
\draw (0,0)[red,fill=red!30] circle (0.4cm);
\draw (0,-0.8) node {\footnotesize{$2$}};
\draw (0,0) node {\footnotesize{$z_0$}};
\draw (2.8,0) circle (0.4cm);
\draw (2.8,-0.8) node {\footnotesize{$2$}};
\draw (2.8,0) node {\footnotesize{$w_2$}};
\draw (0.4,0) -- (1,0);
\draw (1.8,0) -- (2.4,0);
\draw (1.4,0) circle (0.4cm);
\draw (1.4,-0.8) node {\footnotesize{$2$}};
\draw (1.4,0) node {\footnotesize{$w_1$}};
\draw (-0.26,0.3) -- (-0.87,0.87);
\draw (-1.1,1.2) circle (0.4cm);
\draw (-1.1,2) node {\footnotesize{$1$}};
\draw (-1.1,1.2) node {\footnotesize{$z_1$}};
\draw (0.26,0.3) -- (0.87,0.87);
\draw (1.1,1.2) circle (0.4cm);
\draw (1.1,2) node {\footnotesize{$1$}};
\draw (1.1,1.2) node {\footnotesize{$z_3$}};
\draw (0,1.2) node {\footnotesize{$\dots$}};
\draw [decorate,decoration={brace,amplitude=6pt}] (-1.1,2.3) to (1.1,2.3);
\draw (0,2.9) node {\footnotesize{$3$}};
\draw (2.6,0.34) -- (2.3,0.8);
\draw (2.2,1.2) circle (0.4cm);
\draw (3,0.34) -- (3.3,0.8);
\draw (3.4,1.2) circle (0.4cm);
\draw (2.2,1.2) node {\footnotesize{$w_3$}};
\draw (2.2,2) node {\footnotesize{$1$}};
\draw (3.4,2) node {\footnotesize{$1$}};
\draw (3.4,1.2) node {\footnotesize{$w_4$}};
\end{tikzpicture}
}
\caption{\label{fig:DGen3} $\{\mathcal{P}_{[1^3]}(3)$, $\mathcal{P}_{[1^2]}(2)\}$ Quiver with $SU(2)^{3}\times D_4$ global symmetry, $b=1$, $dim\; \mathcal{M}_C ^\mathbb{H} = 10$.}
\end{figure}
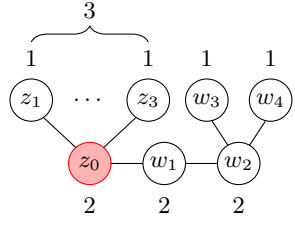

Now, lets study a quiver where we add a rank $2$ node to obtain a chain of two rank $2$ nodes. Furthermore, split the $\mathcal{P}_{[1^5]}(5)$ bouquet into $\mathcal{P}_{[1^3]}(3)$ bouquet attached to the first $2$ node and a $\mathcal{P}_{[1^2]}(2)$ bouquet attached to the other $2$ node. This splitting of nodes justifies the splitting of the fundamental weight fugacities of the bouquet nodes into $x_i$ and $y_i$ in the previous case. The splitting carries over to the HWG such that the highest weight fugacities split into $\mu_i$ and $\nu_i$. The resulting quiver, which now corresponds to a pair of partitions $\{\mathcal{P}_{[1^3]}(3),\mathcal{P}_{[1^2]}(2)\}$, is depicted in Figure \eref{fig:DGen2}.
The balance of the unbalanced red node is $(3*1+2) -(2*2)=1$. One expects ${A_1}^{3} \times D_3 \equiv {A_1}^{3} \times A_3$ global symmetry on the Coulomb branch from simply looking at the balanced sub-quivers. The unrefined HS is computed as
\be
HS(t)= \frac{P_2(t)}{(1-t)^{16} (1+t)^8 (1+t+t^2)^8}
\ee
where
\be
 \begin{split}
P_2(t) &= 1 + 16 t^2 + 40 t^3 + 118 t^4 + 336 t^5 + 747 t^6 + 1344 t^7 + 2396 t^8 +3616 t^9 \\
&+ 4670 t^{10} + 5568 t^{11} + 6060 t^{12} +\cdots palindrome \cdots +t^{24}.
 \end{split}
 \ee
 The expansion of the unrefined HS yields
 \be
HS(t)=1 + 24 t^2 + 48 t^3 + 282 t^4 + 848 t^5 + 2743 t^6 + 7728 t^7 + O(t^8),
 \ee
 and the corresponding unrefined PL takes the form
 \be
 PL=24 t^2 + 48 t^3 - 18 t^4 - 304 t^5 - 601 t^6 + 1488 t^7 + O(t^8).
 \ee
The $t^2$ coefficient in the last equation can be identified with the total dimension of the adjoint representations that form the global symmetry:
\be
3\times dim\;[2]_{A_1} + dim\;[0,1,1]_{D_3} = 24.
\ee
Given the simple root fugacity assignment in Figure \eref{fig:DGen2fugs},
perform a mapping according to Equations \eref{FM1} and \eref{FM2}
 \begin{gather}
 z_1\rightarrow x_1^2,\; z_2\rightarrow x_2^2,\; z_3\rightarrow x_3^2, \label{FM1} \\ 
 z_6 \rightarrow y_1^2 y_2^{-1} y_3^{-1},\; z_5 \rightarrow y_2^2 y_1^{-1}, \; z_7 \rightarrow y_3^2 y_1^{-1} \label{FM2} \\ 
 z_4\rightarrow (z_1\; z_2 \;z_3 \;{z_5}^2 \;z_6 \;z_7)^{-\frac{1}{2}}= (x_1^{-2} x_2^{-2} x_3^{-2} y_1 y_2^{-3} y_3^{-1})^{\frac{1}{2}},  \label{FM3}
 \end{gather}
 where $x_i,\; i=1,2,3$ are the $SU(2)$ fundamental weight fugacities, $y_i,\; i=1,2,3$ are the $D_3$ fundamental weight fugacities, and $z_4$, the fugacity of the red node is eliminated according to Equation \eref{FM3}. The resulting HWG takes the form given by Equation \eqref{eq:G2}.

\begin{equation} \label{eq:G2}
HWG= PE[ \left(\nu_1 ^2 +\nu_2^2+\nu_3^2 + \mu_2\mu_3\right)t^2 + (\nu_1\nu_2\nu_3\mu_1)(t^3+t^5) + \mu_1^2t^4 +t^4 - (\nu_1\nu_2\nu_3\mu_1)^2 t^{10}]
\end{equation}
Note the slight change in the structure of the terms appearing in Equation \eqref{eq:G2} compared to Equation \eqref{eq:G1}. At order $t^2$ there is the adjoint $[0,1,1]$ rep of $D_3$ and the three-adjoint $[2;2;2]$ rep of $SU(2)^3$. Recall that $[d_1,d_2,d_3]$ and $[a_1;a_2;a_3]$ denote the Dynkin labels of $D_3$ and $SU(2) \times SU(2) \times SU(2)$, respectively. At $t^3$ and $t^5$ there are generators transforming under  $[1;1;1;1,0,0]$ representation of $SU(2)^3 \times D_3$. There are also generators transforming under $[0;0;0;0,0,0]$ and $[0;0;0;2,0,0]$ at $t^4$. Finally, there is a relation at order $t^{10}$ transforming under $[2;2;2;2,0,0]$. Before the identification of a general pattern of HWG for this family of quivers is made possible, one more case needs to be considered. For this purpose, consider the quiver in Figure \eref{fig:DGen3}, with the main chain consisting of three rank $2$ nodes. The simple root fugacities are indicated inside the nodes in Figure \eref{fig:DGen3}.
The anticipated global symmetry, read off as the balanced sub-diagrams, is ${A_1}^3 \times D_4$. After the computation of the HS with the indicated simple root fugacities, the unrefined HS is obtained by setting all the fugacities to unity. The result is given in Equation \eqref{eq:hsD41}
\be \label{eq:hsD41}
HS_{[1^3]}(t)= \frac{P_3(t)}{(-1 + t)^{20} (1 + t)^{12} (1 + t + t^2)^{10}},
   \ee
   where
  \be
  \begin{split}
 Q_1(t)&=  1 + 2 t + 28 t^2 + 108 t^3 + 440 t^4 + 1482 t^5 + 4394 t^6 + 11122 t^7 + 25532 t^8 \\
    &+ 52164 t^9 + 95692 t^{10} + 158586 t^{11} + 239637 t^{12} + 328584 t^{13} + 410844 t^{14} \\
     &+ 469872 t^{15} + 491976 t^{16} + \cdots palindrome \cdots + t^{32}.
   \end{split}
   \ee
The expansion of the unrefined HS has the form
\be
HS(t)=1 + 37 t^2 + 64 t^3 + 630 t^4 + 1728 t^5 + 7803 t^6 + 22848 t^7 + 
 75858 t^8 +O(t^9).
 \ee
 Taking the PL of Equation \eref{eq:hsD41} one finds
 \begin{equation} \label{eq:plD4-1}
PL= 37t^2 + 64 t^3 - 73 t^4 - 640 t^5 - 715 t^6 + 6208 t^7 + 23614 t^8 - O(t^9) .
\end{equation}
The $t^2$ coefficient in Equation \eref{eq:plD4-1} matches the dimension of the adjoint representation of the global symmetry:
 \be
 3\times dim\; [2]_{A_1} + dim\;[0,1,0,0]_{D_4} = 37.
 \ee
Let us use the fugacity map 
 \begin{gather}
 z_i\rightarrow x_i^2,\; i = 1,2,3 \label{eq:FM21} \\
 w_1 \rightarrow y_1^2 y_2^{-1},\; w_2 \rightarrow y_2^2 y_1^{-1} y_3^{-1} y_4^{-1}, \; w_3 \rightarrow y_3^2 y_2^{-1}, \; w_4 \rightarrow y_4^2 y_2^{-1} \label{eq:FM22}\\
 z_0\rightarrow (z_1 z_2 z_3  w_1^2 w_2^2 w_3 w_4)^{-\frac{1}{2}} = (x_1 x_2 x_3 y_1)^{-1}. \label{eq:FM23}
 \end{gather}
whereupon the simple root fugacities $z_i,\; i=1,2,3$, and $w_j,\; j=1,2,3,4$, map to the fudamental weight fugacities of the $SU(2)$ and $D_4$, respectively, and $z_0$, the fugacity of the red node, is substituted according to prescription \eref{eq:FM23}. The HWG takes the form given by Equation \eref{eq:G3}
\begin{equation} \label{eq:G3}
HWG= PE[ \left(\nu_1 ^2 +\nu_2^2+\nu_3^2 + \mu_2\right)t^2 + (\nu_1\nu_2\nu_3\mu_1)(t^3+t^5) + \mu_1^2 t^4 +t^4 - (\nu_1\nu_2\nu_3\mu_1)^2 t^{10}],
\end{equation}
where, $\mu_2$ is the adjoint weight fugacity of $D_4$. Recall that the vector reps appear because the vector nodes of the balanced Dynkin sub-diagrams connect to the unbalanced node. Moreover, these appear at orders $t^3$ and $t^5$ since the imbalance is $1$, same as in the previous case. This form of HWG is anticipated for all members of this family. 

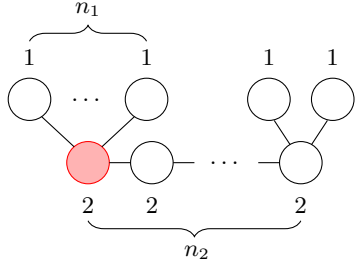
\begin{figure}[h!]
\center{
\begin{tikzpicture}[scale=0.70]
\draw (0,0)[red,fill=red!30] circle (0.4cm);
\draw (0,-0.8) node {\footnotesize{$2$}};
\draw (1.2,0) circle (0.4cm);
\draw (1.2,-0.8) node {\footnotesize{$2$}};
\draw (4,0) circle (0.4cm);
\draw (4,-0.8) node {\footnotesize{$2$}};
\draw (0.4,0) -- (0.8,0);
\draw (3.2,0) -- (3.6,0);
\draw (1.6,0) -- (2,0);
\draw (2.6,0) node {\footnotesize{$\dots$}};
\draw (-0.26,0.3) -- (-0.87,0.87);
\draw (-1.1,1.2) circle (0.4cm);
\draw (-1.1,2) node {\footnotesize{$1$}};
\draw (0.26,0.3) -- (0.87,0.87);
\draw (1.1,1.2) circle (0.4cm);
\draw (1.1,2) node {\footnotesize{$1$}};
\draw (0,1.2) node {\footnotesize{$\dots$}};
\draw [decorate,decoration={brace,amplitude=6pt}] (-1.1,2.3) to (1.1,2.3);
\draw (0,2.9) node {\footnotesize{$n_1$}};
\draw (3.8,0.34) -- (3.5,0.8);
\draw (3.4,1.2) circle (0.4cm);
\draw (4.2,0.34) -- (4.5,0.8);
\draw (4.6,1.2) circle (0.4cm);
\draw (3.4,2) node {\footnotesize{$1$}};
\draw (4.6,2) node {\footnotesize{$1$}};
\draw [decorate,decoration={brace,amplitude=6pt}] (4,-1.1) to (0,-1.1);
\draw (2.05,-1.7) node {\footnotesize{$n_2$}};
\end{tikzpicture}
}
\caption{\label{fig:DGenGen} $\{\mathcal{P}_{[1^{n_1}]}(n_1)$, $\mathcal{P}_{[1^2]}(2)\}$ Quiver with $SU(2)^{n_1}\times D_{n_2+1}$ global symmetry, $b=n_1 +2 -4$, $dim\;\mathcal{M}_C ^\mathbb{H} = n_1 +2n_2 +1$.}
\end{figure}

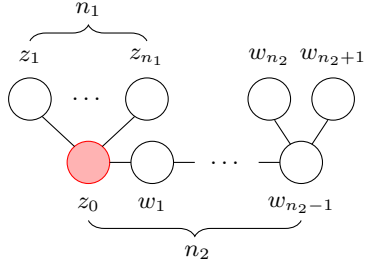
\begin{figure}[h!]
\center{
\begin{tikzpicture}[scale=0.70]
\draw (0,0)[red,fill=red!30] circle (0.4cm);
\draw (0,-0.8) node {\footnotesize{$z_0$}};
\draw (1.2,0) circle (0.4cm);
\draw (1.2,-0.8) node {\footnotesize{$w_1$}};
\draw (4,0) circle (0.4cm);
\draw (4,-0.8) node {\footnotesize{$w_{n_2-1}$}};
\draw (0.4,0) -- (0.8,0);
\draw (3.2,0) -- (3.6,0);
\draw (1.6,0) -- (2,0);
\draw (2.6,0) node {\footnotesize{$\dots$}};
\draw (-0.26,0.3) -- (-0.87,0.87);
\draw (-1.1,1.2) circle (0.4cm);
\draw (-1.1,2) node {\footnotesize{$z_1$}};
\draw (0.26,0.3) -- (0.87,0.87);
\draw (1.1,1.2) circle (0.4cm);
\draw (1.1,2) node {\footnotesize{$z_{n_1}$}};
\draw (0,1.2) node {\footnotesize{$\dots$}};
\draw [decorate,decoration={brace,amplitude=6pt}] (-1.1,2.3) to (1.1,2.3);
\draw (0,2.9) node {\footnotesize{$n_1$}};
\draw (3.8,0.34) -- (3.5,0.8);
\draw (3.4,1.2) circle (0.4cm);
\draw (4.2,0.34) -- (4.5,0.8);
\draw (4.6,1.2) circle (0.4cm);
\draw (3.4,2) node {\footnotesize{$w_{n_2}$}};
\draw (4.6,2) node {\footnotesize{$w_{n_2+1}$}};
\draw [decorate,decoration={brace,amplitude=6pt}] (4,-1.1) to (0,-1.1);
\draw (2.05,-1.7) node {\footnotesize{$n_2$}};
\end{tikzpicture}
}
\caption{\label{fig:DGenGenFug} $\{\mathcal{P}_{[1^{n_1}]}(n_1)$, $\mathcal{P}_{[1^2]}(2)\}$ Quiver with fugacity assignment.}
\end{figure}

Increasing the number of nodes of the left bouquet results in a simple change of the form of the HWG. The number of rank $1$ nodes in the left bouquet only changes the imbalance of the red node, therefore, for higher $n_1$, the $(t^3+t^5)$ terms will appear at higher orders of $t$ \cite{MU18}. For the $\{\mathcal{P}_{[1^4]}(4)$, $\mathcal{P}_{[1^2]}(2)\}$ theory one expects the contribution to appear at $(t^4+t^6)$. More generally, for a $\{\mathcal{P}_{[1^{n_1}]}(n_1)$, $\mathcal{P}_{[1^2]}(2)\}$ theory, these contributions are expected at orders $(t^{n_1} +t^{n_1+2})$. We are in a position to write down a general expression of the HWG for a two parameter family of bouquet quivers of the form in Figure \eref{fig:DGenGen}. 
The simple root fugacity assignment is shown separately in Figure \eref{fig:DGenGenFug} for clarity of presentation.
Perform the mapping such that the simple root fugacities $z_i$, $i=1,\dots, n_1$, map to the $SU(2)$ fundamental weight fugacities $\nu_i$, $i=1,\dots, n_1$ and, the root fugacities $w_j,\; j=1,\dots,n_2+1$, map to $D_{n_2+1}$ fundamental weight fugacities $\mu_j$, $j=1,\dots, n_2+1$. In full analogy to previous cases, this is achieved by deriving the fugacity map using the Cartan matrix, and the elimination of the $z_0$ fugacity of the red node follows from the gauge fixing condition \eref{eq:constraint1}. The HWG for the family of $\{\mathcal{P}_{[1^{n_1}]}(n_1)$, $\mathcal{P}_{[1^2]}(2)\}$ quiver theories takes the form given by the general Formula \eref{eq:Gen}.
\begin{equation}\label{eq:Gen}
HWG= PE[ (\sum_{i=1}^{n_1} \nu_i ^2+ \mu_2)t^2 + (\mu_1 \prod_{i=1}^{n_1} \nu_i )(t^{n_1}+t^{n_1+2}) + (\mu_1^2)t^4 +t^4 - (\mu_1 \prod_{i=1}^{n_1} \nu_i)^2 t^{10}]
\end{equation}
Formula \eref{eq:Gen} contains the usual generators at order $t^2$ transforming under the adjoint representations corresponding to the $D_{n_2 +1}$ and $SU(2)$ nodes. In addition, there are generators in the vector representation of $D_{n_2 +1}$ since it is the vector Dynkin node that connects to the unbalanced node. For the same reason, the ($n_1$)-fundamental representation of the $SU(2)$ is present. Formula \eref{eq:Gen} is verified with an explicit computation of the HWG up to $n_1=4$ and $n_2=4$. \\

General analogue of Formula \eref{eq:Gen} for bouquet quivers with ABCEFG factors in the global symmetry is conjectured in section \ref{5}. 

\subsection{Discrete Gauging of $\{ \mathcal{P}_{[1^3]}(3),n_2 = 3\}$ theory}

In this subsection, we use the unrefined description of the Coulomb branches to provide a non-trivial test of Equation \eref{eq:00}. Let us study the two parameter family of theories in Figure \eref{fig:DGenGen}. Note, that the quiver has discrete global symmetries $S_{n_1}$ and $S_2$ corresponding to the left and the right bouquet, respectively. Lets focus on the first bouquet only, such that the rest of the quiver preserves a manifest $D_{n_2+1}$ global symmetry. Such theories will be denoted by $\{ \mathcal{P}_{[n_1]}(n_1),n_2\}$. The two parameters $n_1$ and $n_2$ correspond to the number of bouquet nodes and the number of rank $2$ chain nodes, respectively. Let us study discrete gauging on a particular member of this family by setting $n_1=n_2=3$. The considered theory is depicted in Figure \eref{fig:DGen3}. The unrefined HS is given by Equation \eref{eq:hsD41} in the previous subsection. 

\subsubsection{Gauging $H_{\lambda}=\mathbb{Z}_2$}

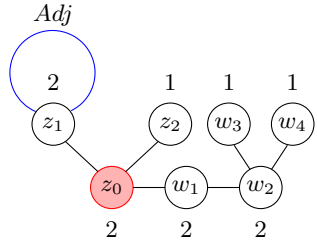
\begin{figure}[h!]
\center{
\begin{tikzpicture}[scale=0.70]
\draw (0,0)[red,fill=red!30] circle (0.4cm);
\draw (0,-0.8) node {\footnotesize{$2$}};
\draw (0,0) node {\footnotesize{$z_0$}};
\draw (2.8,0) circle (0.4cm);
\draw (2.8,-0.8) node {\footnotesize{$2$}};
\draw (2.8,0) node {\footnotesize{$w_2$}};
\draw (0.4,0) -- (1,0);
\draw (1.8,0) -- (2.4,0);
\draw (1.4,0) circle (0.4cm);
\draw (1.4,-0.8) node {\footnotesize{$2$}};
\draw (1.4,0) node {\footnotesize{$w_1$}};
\draw (-0.26,0.3) -- (-0.87,0.87);
\draw (-1.1,1.2) circle (0.4cm);
\draw (-1.1,2) node {\footnotesize{$2$}};
\draw (-1.1,1.2) node {\footnotesize{$z_1$}};
\draw (0.26,0.3) -- (0.87,0.87);
\draw (1.1,1.2) circle (0.4cm);
\draw (1.1,2) node {\footnotesize{$1$}};
\draw (1.1,1.2) node {\footnotesize{$z_2$}};
\draw (2.6,0.34) -- (2.3,0.8);
\draw (2.2,1.2) circle (0.4cm);
\draw (3,0.34) -- (3.3,0.8);
\draw (3.4,1.2) circle (0.4cm);
\draw (2.2,1.2) node {\footnotesize{$w_3$}};
\draw (2.2,2) node {\footnotesize{$1$}};
\draw (3.4,2) node {\footnotesize{$1$}};
\draw (3.4,1.2) node {\footnotesize{$w_4$}};
 \draw[-,blue] (-0.765,1.455) arc (-64:247:0.8);
 \draw (-1.1,3.27) node {\footnotesize{$Adj$}};
\end{tikzpicture}
}
\caption{\label{fig:DC1} $\{ \mathcal{P}_{[2,1]}(3),n_2 = 3\}$ Quiver with $SU(2)^{2}\times D_4$ global symmetry, $b=1$, $dim\; \mathcal{M}_C ^\mathbb{H} = 10$.}
\end{figure}

 Gauge a $\mathbb{Z}_2$ subgroup of the discrete global $S_3$ symmetry to construct a new theory. Following Conjecture \eref{Main Conjecture} one obtains the quiver depicted in Figure \eref{fig:DC1}, which is accordingly denoted by $\{ \mathcal{P}_{[2,1]}(3),3\}$. The balanced subset of the quiver forms the ${A_1}^2 \times D_4$ Dynkin diagrams. Hence, the anticipated global symmetry is $SU(2) \times SU(2) \times SO(8)$.
The unrefined HS is given by Equation \eref{eq:hsD42}
\be \label{eq:hsD42}
HS_{[2,1]}(t)= \frac{Q_2(t)}{(-1 + t)^{20} (1 + t)^{14} (1 + t^2)^2 (1 + t + t^2)^{10}},
 \ee
 where
 \be
  \begin{split}
 Q_2(t)&= 1 + 4 t + 32 t^2 + 146 t^3 + 592 t^4 + 2052 t^5 + 6348 t^6 + 17276 t^7 + 42495 t^8  \\ 
 &+94722 t^9 +192829 t^{10} + 359694 t^{11} + 618737 t^{12} + 983550 t^{13} + 1449871 t^{14}  \\ 
 &+1985584 t^{15} + 2531833 t^{16} + 3008328 t^{17} + 3335694 t^{18}  \\
  &+3452040 t^{19}+ \cdots palindrome \cdots + t^{38}.
   \end{split}
   \ee
From the unrefined PL  given by Equation \eref{eq:plD4-2},
  \begin{equation} \label{eq:plD4-2}
PL=34 t^2 + 48 t^3 - 66 t^4 - 400 t^5 - 129 t^6 + 3744 t^7 + 7875 t^8 - 
 28352 t^9+O(t^{10})
\end{equation}
observe that the $t^2$ coefficient agrees with the expected global symmetry:
\be
2\times dim [2]_{A_1} + dim [0,1,0,0]_{D_4} = 34.
\ee

\subsubsection{Gauging $H_{\lambda}=S_3$}

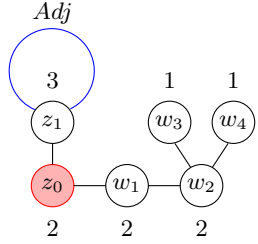
\begin{figure}[h!]
\center{
\begin{tikzpicture}[scale=0.70]
\draw (0,0)[red,fill=red!30] circle (0.4cm);
\draw (0,-0.8) node {\footnotesize{$2$}};
\draw (0,0) node {\footnotesize{$z_0$}};
\draw (2.8,0) circle (0.4cm);
\draw (2.8,-0.8) node {\footnotesize{$2$}};
\draw (2.8,0) node {\footnotesize{$w_2$}};
\draw (0.4,0) -- (1,0);
\draw (1.8,0) -- (2.4,0);
\draw (1.4,0) circle (0.4cm);
\draw (1.4,-0.8) node {\footnotesize{$2$}};
\draw (1.4,0) node {\footnotesize{$w_1$}};
\draw (0,1.2) circle (0.4cm);
\draw (0,2) node {\footnotesize{$3$}};
\draw (0,1.2) node {\footnotesize{$z_1$}};
\draw (0,0.4) -- (0,0.8);
\draw (2.6,0.34) -- (2.3,0.8);
\draw (2.2,1.2) circle (0.4cm);
\draw (3,0.34) -- (3.3,0.8);
\draw (3.4,1.2) circle (0.4cm);
\draw (2.2,1.2) node {\footnotesize{$w_3$}};
\draw (2.2,2) node {\footnotesize{$1$}};
\draw (3.4,2) node {\footnotesize{$1$}};
\draw (3.4,1.2) node {\footnotesize{$w_4$}};
 \draw[-,blue] (0.335,1.455) arc (-64:247:0.8);
 \draw (0,3.27) node {\footnotesize{$Adj$}};
\end{tikzpicture}
}
\caption{\label{fig:DC2} $\{ \mathcal{P}_{[3]}(3),n_2 = 3\}$ Quiver with $SU(2) \times D_4$ global symmetry, $b=1$, $dim\; \mathcal{M}_C ^\mathbb{H} = 10$.}
\end{figure}

Finally, gauge $H = S_3$, the entire global symmetry of the $\{ \mathcal{P}_{[1^3]}(3),n_2 = 3\}$ theory to obtain the $\{ \mathcal{P}_{[3]}(3),n_2 = 3\}$ theory depicted in Figure \eref{fig:DC2}. The expected global symmetry is $A_1 \times D_4$.
The unrefined HS is given by Equation \eref{eq:hsD43}
\be \label{eq:hsD43}
HS_{[3]}(t)= \frac{Q_3(t)}{(-1 + t)^{20} (1 + t)^{14} (1 + t^2)^2 (1 + t + t^2)^{10}},
\ee
where
 \be
  \begin{split}
 Q_3(t)&=  1 + 4 t + 29 t^2 + 118 t^3 + 436 t^4 + 1342 t^5 + 3754 t^6 + 9232 t^7 + 20764 t^8  \\
 &+ 42590 t^9 + 80758 t^{10} + 141402 t^{11} + 230675 t^{12} + 350568 t^{13} + 498471 t^{14}  \\ 
  &+663084 t^{15} + 827454 t^{16} + 968184 t^{17} + 1064154 t^{18} + 1097832 t^{19} \\
 &+ \cdots palindrome \cdots + t^{32}.
   \end{split}
   \ee
 The unrefined PL takes the form given by Equation \eref{eq:plD4-3},
 \begin{equation} \label{eq:plD4-3}
PL= 31 t^2 +32 t^3 - 65 t^4 - 224 t^5 + 249 t^6 + 2192 t^7 + 22 t^8 - 21600 t^9+O(t^{10}),
\end{equation}
which is in agreement with the expectation of the global symmetry since the $t^2$ coefficient equals
\be
dim [2]_{A_1} + dim [0,1,0,0]_{D_4} = 31.
\ee

 \subsubsection{Comparison of the Coulomb branch volumes}
 
 Expanding the unrefined Hilbert series \eref{eq:hsD41}, \eref{eq:hsD42} and \eref{eq:hsD43} according to Equation \eref{eq:expansionHS} and plugging into \eref{eq:01} one finds:
 \begin{gather}
 \frac{vol(\CC_{[1^3]})}{vol(\CC_{[2,1]})} = \frac{R_{[1^3]}}{R_{[2,1]}} = \frac{\frac{56791}{3359232}} {\frac{56791}{6718464}} = 2 = ord(\mathbb{Z}_2) \label{eq:R1}  \\
  \frac{vol(\CC_{[1^3]})}{vol(\CC_{[3]})} = \frac{R_{[1^3]}}{R_{[3]}} = \frac{\frac{56791}{3359232}} {\frac{56791}{20155392}} = 6 = ord(S_3) \label{eq:R2} 
 \end{gather}
 
 which are the expected ratios. Obtained results \eref{eq:R1} and \eref{eq:R2} are in accord with Conjecture \eref{Main Conjecture} and provide a non-trivial check that the Coulomb branches of $\{ \mathcal{P}_{[2,1]}(3),n_2 = 2\}$ and $\{ \mathcal{P}_{[3]}(3),n_2 = 2\}$ quivers are $\mathbb{Z}_2$ and $S_3$ orbifolds of the parent $\{ \mathcal{P}_{[1^3]}(3),n_2 = 2\}$ Coulomb branch, respectively. Note, that it follows that the $\{ \mathcal{P}_{[3]}(3),n_2 = 2\}$ Coulomb branch is a $\mathbb{Z}_3$ orbifold of the $\{ \mathcal{P}_{[2,1]}(3),n_2 = 2\}$ Coulomb branch. Again, the explicit test involves the utilization of the methods of the stepwise projection \cite{Stepwise}. Let us now we turn to the third family of quivers.
 
\section{Third Family: $A$-type Bouquet Quivers with $U(1)^n \times {A_2}^2$ global symmetry} \label{3}

\begin{figure}[h!]
\center{
\begin{tikzpicture}[scale=0.70]
\draw (0.4,0) -- (0.8,0);
\draw (0,0)[red,fill=red!30] circle (0.4cm);
\draw (0,-0.8) node {\footnotesize{$3$}};
\draw (1.2,-0.8) node {\footnotesize{$2$}};
\draw (2.4,0) circle (0.4cm);
\draw (2.4,-0.8) node {\footnotesize{$1$}};
\draw(1.2,0) circle (0.4cm);
\draw (-2.4,0) circle (0.4cm);
\draw (-2.4,-0.8) node {\footnotesize{$1$}};
\draw (-0.4,0) -- (-0.8,0);
\draw (1.6,0) -- (2,0);
\draw (-1.2,0) circle (0.4cm);
\draw (-1.2,-0.8) node {\footnotesize{$2$}};
\draw (-1.6,0) -- (-2,0);
\draw (-0.26,0.3) -- (-0.87,0.87);
\draw (-1.1,1.2)[red,fill=red!30] circle (0.4cm);
\draw (-1.1,2) node {\footnotesize{$1$}};
\draw (0.26,0.3) -- (0.87,0.87);
\draw (1.1,1.2)[red,fill=red!30] circle (0.4cm);
\draw (1.1,2) node {\footnotesize{$1$}};
\draw (0,0.4) -- (0,0.8);
\draw (0,1.2)[red,fill=red!30] circle (0.4cm);
\draw (0,2) node {\footnotesize{$1$}};
\end{tikzpicture}
}
\caption{\label{fig:Impl1} $\mathcal{P}_{[1^{3}]}(3)$ Quiver with $SU(2)^2 \times U(1)^3$ global symmetry, $b_i=1,\; i=1,2,3,4$, $dim\;\mathcal{M}_C ^\mathbb{H} = 11$.}
\end{figure}
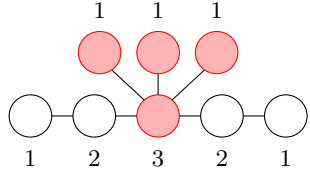

In this section, we consider unitary bouquet quivers with $U(1)^n \times {A_{k-1}}^2$ global symmetry, following the parametrization in Figure \eref{fig:4}. Let us set $k=n=3$ to obtain the theory depicted in Figure \eref{fig:Impl1}. As a gauge theory, the Coulomb branch quivers in this section correspond to the Higgs branches\footnote{Note, that herein, when we talk about the $U(1)$ factors in the global symmetry, we are referring to the 3d quivers only. It should be emphasized that in the 6d, the anomalous $U(1)$ factors are no longer part of the global symmetry. Nevertheless, they remain as part of the isometry of the moduli space.} of quivers describing a 6d $\CN=(1,0)$ low energy dynamics of a stack of three M5 branes on an $A_2$ singularity \cite{HZ18}. The arrangement of the bouquet nodes corresponds to three separated $M5$ branes, hence, we accordingly denote this theory by $\mathcal{P}_{[1^3]}(3)$. In the previous sections, only \emph{fully balanced} or \emph{minimally unbalanced} quivers were considered. Since in this section, we encounter quivers with more than one unbalanced node, the conjectured prescription for reading off the global symmetry from the quiver needs to be extended. For the global symmetry of a quiver with $N$, $N\geq 2$ unbalanced nodes, there holds
\be
G_{global} = G^{\;i}_{balanced}\times U(1)^{N-1},
\ee
where $G^{\;i}_{balanced}$ is the symmetry group that corresponds to the Dynkin diagram formed by the $i$-th balanced subset of nodes.  Moreover, there are $N-1$ additional $U(1)$ factors such that the number of $U(1)$ Abelian factors in the global symmetry is one less than the number of unbalanced nodes. In the case of Figure \eref{fig:Impl1}, one expects $4-1=3$ copies of such Abelian factors. Hence, the expected global symmetry is $U(1)^3 \times SU(3)^2$. There is an additional $S_3$ discrete global symmetry that permutes the bouquet nodes. The balance of all four unbalanced nodes is $1$. Throughout this section, we refrain from showing explicit fugacity assignments and maps since the main objective is to test Formula \eref{eq:00}.
The unrefined Hilbert Series for the $\mathcal{P}_{[1^3]}(3)$ quiver takes the form
\begin{equation} \label{eq:hsA1}
HS_{[1^3]}(t)=\frac{P_1(t)}{(-1+t)^{22} (1+t)^{16} (1+t^2)^8 (1+t+t^2)^{11} (1+t+t^2+ 
     t^3 + t^4)^5},
\end{equation}
where
 \be
  \begin{split}
P_1 (t)&=1 + 10 t + 66 t^2 + 343 t^3 + 1561 t^4 + 6421 t^5 + 24318 t^6 + 
 85373 t^7 + 279505 t^8\\
 & + 856911 t^9 +2470009 t^{10} + 6715986 t^{11} + 17278135 t^{12} + 42171723 t^{13} + 97892626 t^{14}  \\
 &+  216588291 t^{15} + 457659547 t^{16} + 925229636 t^{17} + 1792503575 t^{18} + \\
 &+3332789141 t^{19} + 5954799253 t^{20} + 10236605469 t^{21}  \\
 &+16948970150 t^{22} + 27055291005 t^{23} + 41673945980 t^{24}  \\
 &+61990354851 t^{25} + 89112653186 t^{26} + 123875740431 t^{27}  \\
 &+166613606315 t^{28} + 216934711187 t^{29} + 273547259468 t^{30}  \\
 &+334183688804 t^{31} + 395665660521 t^{32} + 454128806740 t^{33}  \\
 &+505396609910 t^{34} + 545458043162 t^{35} + 570976321490 t^{36}  \\
 &+579740398924 t^{37} + \cdots palindrome \dots + t^{74}. 
\end{split}
\ee
Taking the PL of the unrefined HS yields
\be
PL=19t^2+24t^3+53t^4+36t^5-129 t^6-588t^7- \\
1347t^8-O(t^9).
\ee
The $t^2$ coefficient agrees with the anticipated $G_{global}=U(1)^3 \times SU(3)^2$ since
\be
3 \times dim\; U(1) + 2 \times dim\; [1,1]_{A_2} = 19.
\ee

\subsubsection{Gauging $H_{\lambda}=\mathbb{Z}_2$}

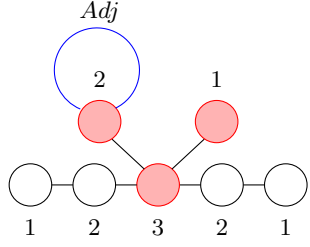
\begin{figure}[h!]
\center{
\begin{tikzpicture}[scale=0.70]
\draw (0.4,0) -- (0.8,0);
\draw (0,0)[red,fill=red!30] circle (0.4cm);
\draw (0,-0.8) node {\footnotesize{$3$}};
\draw (1.2,-0.8) node {\footnotesize{$2$}};
\draw (2.4,0) circle (0.4cm);
\draw (2.4,-0.8) node {\footnotesize{$1$}};
\draw(1.2,0) circle (0.4cm);
\draw (-2.4,0) circle (0.4cm);
\draw (-2.4,-0.8) node {\footnotesize{$1$}};
\draw (-0.4,0) -- (-0.8,0);
\draw (1.6,0) -- (2,0);
\draw (-1.2,0) circle (0.4cm);
\draw (-1.2,-0.8) node {\footnotesize{$2$}};
\draw (-1.6,0) -- (-2,0);
\draw (-0.26,0.3) -- (-0.87,0.87);
\draw (-1.1,1.2)[red,fill=red!30] circle (0.4cm);
\draw (-1.1,2) node {\footnotesize{$2$}};
\draw (0.26,0.3) -- (0.87,0.87);
\draw (1.1,1.2)[red,fill=red!30] circle (0.4cm);
\draw (1.1,2) node {\footnotesize{$1$}};
  \draw[-,blue] (-0.8,1.455) arc (-64:247:0.8);
  \draw (-1.1,3.27) node {\footnotesize{$Adj$}};
\end{tikzpicture}
}
\caption{\label{fig:Impl2} $\mathcal{P}_{[2,1]}(3)$ Quiver with $SU(3)^2 \times SU(2)^2 \times U(1)^2$ global symmetry, $dim\;\mathcal{M}_C ^\mathbb{H} = 11$.}
\end{figure}

Gauge a $\mathbb{Z}_2$ subgroup of the discrete $S_3$ symmetry such that, according to Conjecture \eref{Main Conjecture}, the obtained theory corresponding to $\mathcal{P}_{[2,1]}(3)$ is described by a quiver in Figure \eref{fig:Impl2}. The two balanced sub-quivers form a $A_2 \times A_2$ global symmetry. Moreover, there are three unbalanced nodes which implies that there are two additional $U(1)$ factors in the global symmetry. Altogether, we have
\be
G_{global} = U(1)^2 \times SU(3)^2.
\ee 
The unrefined Hilbert Series is given by Equation \eref{eq:hsA2},
\begin{equation} \label{eq:hsA2}
HS_{[2,1]}(t)=\frac{P_2(t)}{(-1+t)^{22} (1+t)^{16} (1+t^2)^8 (1-t+t^2) (1+t+t^2)^{11} (1+t+t^2+ 
     t^3 + t^4)^5},
\end{equation}
where
 \be
  \begin{split}
P_2 (t)&=1 + 9 t + 56 t^2 + 276 t^3 + 1192 t^4 + 4635 t^5 + 16581 t^6 + 55030 t^7 + 170775 t^8 \\
 &+ 497861 t^9 + 1369519 t^{10} + 3566403 t^{11} + 8819153 t^{12} +20761818 t^{13} \\
 &+ 46641268 t^{14} + 100192056 t^{15} + 206191600 t^{16} + 407200034 t^{17} \\
 &+ 772867324 t^{18} + 1411740354 t^{19} +2484834652 t^{20}+ 4219097138 t^{21}\\
 & + 6917735891 t^{22} + 10963035811 t^{23} + 16806739624 t^{24} + 24943050628 t^{25}  \\
&+ 35861261184 t^{26} + 49977850045 t^{27} + 67552995501 t^{28} + 88601153016 t^{29}\\
& + 112810770236 t^{30} + 139490143344 t^{31} +167556757817 t^{32} + 195581752669 t^{33}\\
&+ 221893836645 t^{34} + 244734679875 t^{35} + 262447986225 t^{36} + 273674136136 t^{37}  \\
 &+277519995798 t^{38} + \cdots palindrome \dots + t^{76}.
\end{split}
\ee
The unrefined PL is given by Equation \eref{eq:plA1}.
\be \label{eq:plA1}
PL=18t^2 + 22 t^3 + 36 t^4 + 20 t^5 - 71 t^6 - 320 t^7 - 615 t^8 - 
   O(t^9).
\ee
The $t^2$ coefficient agrees with the anticipation of the global symmetry since
\be
2 \times dim\; U(1) + 2 \times dim\; [1,1]_{A_2} = 18.
\ee

\subsubsection{Gauging $H_{\lambda}=S_3$}

\begin{figure}[h!]
\center{
\begin{tikzpicture}[scale=0.70]
\draw (0.4,0) -- (0.8,0);
\draw (0,0)[red,fill=red!30] circle (0.4cm);
\draw (0,-0.8) node {\footnotesize{$3$}};
\draw (1.2,-0.8) node {\footnotesize{$2$}};
\draw (2.4,0) circle (0.4cm);
\draw (2.4,-0.8) node {\footnotesize{$1$}};
\draw(1.2,0) circle (0.4cm);
\draw (-2.4,0) circle (0.4cm);
\draw (-2.4,-0.8) node {\footnotesize{$1$}};
\draw (-0.4,0) -- (-0.8,0);
\draw (1.6,0) -- (2,0);
\draw (-1.2,0) circle (0.4cm);
\draw (-1.2,-0.8) node {\footnotesize{$2$}};
\draw (-1.6,0) -- (-2,0);
\draw (0,0.4) -- (0,0.8);
\draw (0,1.2)[red,fill=red!30] circle (0.4cm);
\draw (0.8,1.2) node {\footnotesize{$3$}};
  \draw[-,blue] (0.335,1.455) arc (-64:247:0.8);
  \draw (0,3.27) node {\footnotesize{$Adj$}};
\end{tikzpicture}
}
\caption{\label{fig:Impl3} $\mathcal{P}_{[3]}(3)$ Quiver with $SU(3)^2 \times SU(2) \times U(1)$ global symmetry, $b_3 =1$, $b_{Adj} =1$, $dim\;\mathcal{M}_C ^\mathbb{H} = 11$.}
\end{figure}
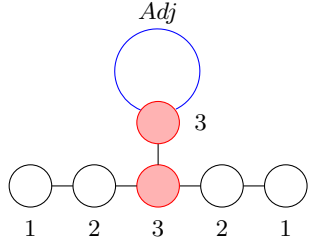

Finally, gauge the entire $S_3$ of $\mathcal{P}_{[1^3]}(3)$ in Figure \eref{fig:Impl1}. According to Conjecture \eref{Main Conjecture}, one obtains the $\mathcal{P}_{[3]}(3)$ quiver, depicted in Figure \eref{fig:Impl3}. Since there are two unbalanced nodes, a single $U(1)$ factor is expected to be present in the global symmetry and we have:
\be
G_{global} = U(1) \times SU(3) \times SU(3)
\ee
Balance of the central and adjoint $3$ node is $b_3=1$ and $b_{Adj}=1$, respectively. 
The lack of the $S_3$ symmetry of the bouquet is reflected in the form of the Hilbert Series \eref{eq:hsA3},
\begin{equation} \label{eq:hsA3}
HS_{[3]}(t)=\frac{P_3(t)}{(-1 + t)^{22} (1 + t)^{16} (1 + t^2)^8 (1 - t + t^2) (1 + t + t^2)^{11} (1 +
    t + t^2 + t^3 + t^4)^5}
\end{equation}
where
 \be
  \begin{split}
P_3 (t)&=1 + 9 t + 55 t^2 + 265 t^3 + 1100 t^4 + 4069 t^5 + 13742 t^6 + 42912 t^7 + 125138 t^8 \\
 &+ 343023 t^9 + 888619 t^{10} + 2184322 t^{11} + 5112353 t^{12} + 11424591 t^{13} + 
  24436388 t^{14} \\
  &+ 50131522 t^{15} + 98823582 t^{16} + 187490947 t^{17} + 342838440 t^{18} \\
  &+ 604970597 t^{19} + 1031345366 t^{20} + 1700334084 t^{21} + 2713413646 t^{22}  \\
 &+4194680213 t^{23} + 6286332847 t^{24} + 9138877284 t^{25} + 12895494665 t^{26} \\
 &+17670886241 t^{27} + 23526392712 t^{28} + 30444409900 t^{29} + 38306534638 t^{30} \\
  &+46880165917 t^{31} + 55818219780 t^{32} + 64674799961 t^{33} + 72937612669 t^{34}  \\
 &+80074444293 t^{35} + 85588479301 t^{36} + 89074448896 t^{37} +90267198678 t^{38} \\
 &+ \cdots palindrome \dots + t^{76}. 
\end{split}
\ee
Equation \eref{eq:plA2} contains the unrefined PL
\be \label{eq:plA2}
PL=17t^2 + 20 t^3 + 18 t^4 + 2 t^5 - 33 t^6 - 122 t^7 - 139 t^8 +
   (t^9),
\ee
and we see that the $t^2$ coefficient matches the dimension of the expected global symmetry:
\be
dim\; U(1) + 2 \times dim\; [1,1]_{A_2} = 17.
\ee
Let us now turn to the comparison of the Coulomb branch volumes.

\subsubsection{Comparison of the Coulomb branch Volumes}

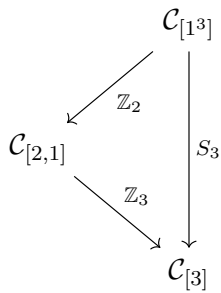
\begin{figure}[h!]
\center{
\begin{tikzcd}[row sep=2.5em]
& \mathcal{C}_{[1^3]} \arrow{dd}{S_3} \arrow{dl}{\mathbb{Z}_2} \\
\mathcal{C}_{[2,1]}  \arrow{dr}{\mathbb{Z}_3} 
& \\ & \mathcal{C}_{[3]}
\end{tikzcd}
}
\caption{\label{fig:Commut3} Commutative diagram of Coulomb branch orbifold hierarchy for $n=3, k=3$ bouquet quivers.}
\end{figure}

 Expanding the unrefined Hilbert series \eref{eq:hsA1}, \eref{eq:hsA2} and \eref{eq:hsA3} according to Equation \eref{eq:expansionHS} and plugging into \eref{eq:01} one finds the ratios:
 \begin{gather}
 \frac{vol(\CC_{[1^3]})}{vol(\CC_{[2,1]})} = \frac{R_{[1^3]}}{R_{[2,1]}} = \frac{\frac{689419303427}{773967052800000}}{\frac{689419303427}{1547934105600000}} = 2 = ord(\mathbb{Z}_2) \label{eq:R1a}  \\
  \frac{vol(\CC_{[1^3]})}{vol(\CC_{[3]})} = \frac{R_{[1^3]}}{R_{[3]}} = \frac{\frac{689419303427}{773967052800000}}{\frac{689419303427}{4643802316800000}} = 6 = ord(S_3) \label{eq:R2a} 
 \end{gather}
Results \eref{eq:R1a} and \eref{eq:R2a} are in accord with Conjecture \eref{Main Conjecture} and they provide a necessary non-trivial check that the Coulomb branches of $\mathcal{P}_{[2,1]}(3)$ and $\mathcal{P}_{[3]}(3)$ are $\mathbb{Z}_2$ and $S_3$ orbifolds of the parent $\mathcal{P}_{[1^3]}(3)$ Coulomb branch, respectively. Note, that the Coulomb branch of $\mathcal{P}_{[3]}(3)$ is a $\mathbb{Z}_3$ quotient of the Coulomb branch of $\mathcal{P}_{[2,1]}(3)$. The orbifold hierarchy for $k=3$, $n=3$ theories is symbolized by the commutative diagram in Figure \eref{fig:Commut3}. Note, that this case is precisely analogical to the cases encountered earlier in this paper (i.e. the $k=2$, $n=1$ quivers of the first family).
In Figure \eref{fig:Commut3}, vertices denote the Coulomb branches of the three $k=3$, $n=3$ $A$-type bouquet quivers and arrows denote the quotients between the branches. The ratios of the Coulomb branch volumes are summarized in Table \eref{tab:Ratios2}. The ratios are in one-to-one correspondence with the ratios of the orders of the corresponding quotient groups. The same analysis was carried out for all members of this family up to $k=4$, $n=5$.
\begin{table}
\begin{center}
\begin{tabular}{ |p{1.8cm}||p{1.6cm}|p{1.6cm}|p{1.6cm}| }
 \hline
 \multicolumn{4}{|c|}{Ratios of $k=3$, $n=3$ Coulomb branch volumes} \\
 \hline
 Partition & $[1^3]$ &$[2,1]$&$[3]$\\
 \hline \hline
 $[1^3]$&1&2&6\\
 $[2,1]$&&1&3\\
$[3]$&&&1\\
 \hline
\end{tabular}
\end{center}
\caption{\label{tab:Ratios2} Ratios of Coulomb branch volumes for $k=3$, $n=3$ family}
\end{table}
Let us now test the discrete gauging construction of Conjecture \eref{Main Conjecture} on the Coulomb branches of non-simply laced quivers.

\newpage

\section{Non-simply Laced Bouquet Quivers} \label{4}

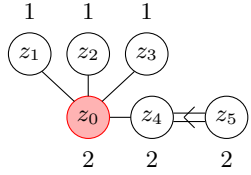
\begin{figure}[h!]
\center{
\begin{tikzpicture}[scale=0.70]
\draw (0.4,0) -- (0.8,0);
\draw (1.6,0.08) -- (2.2,0.08);
\draw (1.6,-0.08) -- (2.2,-0.08);
\draw (1.8,0) -- (2,0.2);
\draw (1.8,0) -- (2,-0.2);
\draw (0,0)[red,fill=red!30] circle (0.4cm);
\draw (0,-0.8) node {\footnotesize{$2$}};
\draw (0,0) node {\footnotesize{$z_0$}};
\draw (1.2,-0.8) node {\footnotesize{$2$}};
\draw (1.2,0) node {\footnotesize{$z_4$}};
\draw (2.6,-0.8) node {\footnotesize{$2$}};
\draw (2.6,0) node {\footnotesize{$z_5$}};
\draw(1.2,0) circle (0.4cm);
\draw(2.6,0) circle (0.4cm);
\draw (-0.26,0.3) -- (-0.87,0.87);
\draw (-1.1,1.2) circle (0.4cm);
\draw (-1.1,2) node {\footnotesize{$1$}};
\draw (-1.1,1.2) node {\footnotesize{$z_1$}};
\draw (0.26,0.3) -- (0.87,0.87);
\draw (1.1,1.2) circle (0.4cm);
\draw (1.1,2) node {\footnotesize{$1$}};
\draw (1.1,1.2) node {\footnotesize{$z_3$}};
\draw (0,0.4) -- (0,0.8);
\draw (0,1.2) circle (0.4cm);
\draw (0,2) node {\footnotesize{$1$}};
\draw (0,1.2) node {\footnotesize{$z_2$}};
\end{tikzpicture}
}
\caption{\label{fig:ImplC21} $\mathcal{P}_{[1^3]}(3)$ Quiver with $SU(2)^3 \times C_2$ global symmetry, $b=1$, $dim\;\mathcal{M}_C ^\mathbb{H} = 8$.}
\end{figure}

This section discusses discrete gauging on Coulomb branches of non-simply laced quiver theories. In a non-simply laced quiver, the non-simply laced edge points towards the short nodes. Accordingly, the side of the quiver that contains short nodes is called \emph{short} and vice versa. Since the discrete gauging action is purely local, it does not distinguish between the \emph{long} and the \emph{short} side of the quiver and, hence, the main construction of this paper can be performed on Coulomb branches of non-simply laced theories in the same fashion as for the simply laced theories. Consider a simple complete bouquet quiver with a $SU(2)^3 \times C_2$ global symmetry. Following the previous notation we denote this theory by $\mathcal{P}_{[1^3]}$. The quiver is depicted in Figure \eref{fig:ImplC21} (the simple root fugacities are shown inside the nodes for completeness). 
One computes the refined HS using the simple root fugacities and sets all to unity to obtain the unrefined Hilbert Series in Equation \eref{eq:1nsl},
\begin{equation}\label{eq:1nsl}
HS_{[1^3]}(t)=\frac{P_4(t)}{(-1 + t)^5 (1 + t)^3 (-1 + t^2)^{10} (1 + t + t^2)^2 (-1 + t^3)^5 (-1 + 
   t^4)^8},
\end{equation}
where
 \be
  \begin{split}
P_4(t)&=1 + 6 t^2 + 25 t^3 + 48 t^4 + 86 t^5 + 174 t^6 - 16 t^7 - 479 t^8 -786 t^9 - 1665 t^{10} \\
  &-2343 t^{11} - 426 t^{12} + 4103 t^{13} + 10658 t^{14} +16406 t^{15} + 9016 t^{16} -13662 t^{17} \\
  &- 35689 t^{18} - 50648 t^{19} -37611 t^{20} + 15375 t^{21} + 69626 t^{22} + 106493 t^{23} \\
&+ 104738 t^{24} + 22330 t^{25} - 97700 t^{26} - 180462 t^{27} - 193479 t^{28} - 82660 t^{29}  \\
  &+109804 t^{30} + 228737 t^{31} + \cdots  palindrome \dots + t^{63} .
 \end{split}
 \ee
 The PL of the unrefined HS is given by Equation \eref{eq:nslPL1}.
 \begin{equation} \label{eq:nslPL1}
PL=19t^2 + 32t^3 + 35t^4 - 64t^5 - 369t^6 - 832 t^7 +O(t^8).
\end{equation}
The $t^2$ coefficient of the unrefined PL is identified as
\be
3 \times dim\;[2]_{A_1} + dim\; [2,0]_{C_2} = 19, 
\ee
which is the dimension of the adjoint representations of the expected global symmetry. 

\subsubsection{Gauging $H_{\lambda}= S_3$}

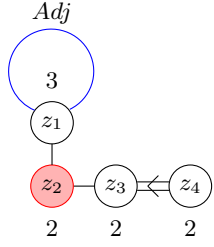
\begin{figure}[h!]
\center{
\begin{tikzpicture}[scale=0.70]
\draw (0.4,0) -- (0.8,0);
\draw (1.6,0.08) -- (2.2,0.08);
\draw (1.6,-0.08) -- (2.2,-0.08);
\draw (1.8,0) -- (2,0.2);
\draw (1.8,0) -- (2,-0.2);
\draw (0,0)[red,fill=red!30] circle (0.4cm);
\draw (0,-0.8) node {\footnotesize{$2$}};
\draw (0,0) node {\footnotesize{$z_2$}};
\draw (1.2,-0.8) node {\footnotesize{$2$}};
\draw (1.2,0) node {\footnotesize{$z_3$}};
\draw (2.6,-0.8) node {\footnotesize{$2$}};
\draw (2.6,0) node {\footnotesize{$z_4$}};
\draw(1.2,0) circle (0.4cm);
\draw(2.6,0) circle (0.4cm);
\draw (0,0.4) -- (0,0.8);
\draw (0,1.2) circle (0.4cm);
\draw (0,2) node {\footnotesize{$3$}};
\draw (0,1.2) node {\footnotesize{$z_1$}};
 \draw[-,blue] (0.335,1.455) arc (-64:247:0.8);
 \draw (0,3.27) node {\footnotesize{$Adj$}};
\end{tikzpicture}
}
\caption{\label{fig:ImplC2Adj} $\mathcal{P}_{[3]}(3)$ Quiver with $A_1 \times C_2$ global symmetry, $b=1$, $dim\;\mathcal{M}_C ^\mathbb{H} = 8$.}
\end{figure}

Let us directly construct the last daughter theory, where the entire $S_3$ discrete global symmetry is gauged. This is achieved by gauging the entire $S_3$ on the Coulomb branch of the $\mathcal{P}_{[1^3]}(3)$ parent quiver. This amounts to the substitution of the original bouquet for a single adjoint $3$ node. Conjecture \eref{Main Conjecture} implies that the resulting quiver (with the root fugacities explicitly indicated inside the nodes for completeness) takes the form depicted in Figure \eref{fig:ImplC2Adj}.
The balanced part of the quiver corresponds to $A_1 \times C_2$ global symmetry. The computation of the unrefined Hilbert Series yields Equation \eref{eq:2nsl}
\begin{equation}\label{eq:2nsl}
HS_{[3]}(t)=\frac{P_5(t)}{(-1 + t)^5 (1 + t)^3 (1 - t^2)^{10} (1 + t^2)^2 (1 + t + t^2)^2 (-1 + 
   t^3)^5 (1 - t^4)^8 (1 + t^2 + t^4)^2},
\end{equation}
where
 \be
  \begin{split}
P_5(t)&=1 + 4 t^2 + 9 t^3 + 26 t^4 + 52 t^5 + 78 t^6 + 34 t^7 - 59 t^8 - 194 t^9 - 454 t^{10} -667 t^{11}\\
& - 911 t^{12} - 918 t^{13} - 48 t^{14} + 1985 t^{15} + 4650 t^{16} + 7296 t^{17}+6956 t^{18}  \\
  & + 1882 t^{19} - 6962 t^{20} - 18740 t^{21} - 25008 t^{22} - 21570 t^{23}-6662 t^{24} \\
&+ 17008 t^{25} + 37396 t^{26} + 47834 t^{27} + 43231 t^{28} + 24580 t^{29} - 6046 t^{30} \\
  &-42257 t^{31} - 77738 t^{32} - 92718 t^{33} - 69502 t^{34} - 11234 t^{35} + 68408 t^{36} \\
  &+ 120258 t^{37} + \cdots  palindrome  \dots + t^{75} . 
 \end{split}
 \ee
  The unrefined PL takes the form:
 \begin{equation}
PL=13t^2 + 16t^3 + 26t^4 +16t^5 - 49t^6 - 264 t^7 +O(t^8).
\end{equation}
Indeed, the $t^2$ coefficient agrees with the dimension of the adjoint representations of the constituent groups of the global symmetry:
\be
dim\;[2]_{A_1} + dim\; [2,0]_{C_2} = 13.
\ee

\subsubsection{Comparison of the Coulomb branch Volumes}

Let us compare the volumes of the two 16 dimensional Coulomb branches computed in this section. The two relevant unrefined Hilbert series are given by Equations \eref{eq:1nsl} and \eref{eq:2nsl}. Expand the HS according to Equation \eref{eq:expansionHS} and plug into \eref{eq:01} to find:
\begin{equation} \label{eq:VolNSL}
\frac{vol(\mathcal{C}_{[1^3]})}{vol(\mathcal{C}_{[3]})}=\frac{\frac{3743}{186624}}{\frac{3743}{1119744}} = 6
\end{equation}
which matches the order of the quotient group $S_3$. The ratio of the volumes of the Coulomb varieties in \eref{eq:VolNSL} provides a non-trivial check that the $\mathcal{C}_{[3]}$ Coulomb branch is a non-Abelian $S_3$ orbifold of the $\mathcal{C}_{[1^3]}$ Coulomb branch. \\


Table \eref{tab:RatiosGEN} lists ratios of the Coulomb branche volumes between a pair of theories of the same type\footnote{By theories of the same type we understand theories given by quivers that differ only in the form of the bouquets.} with bouquets invariant under a discrete $G$ and $H$ symmetry, respectively. 
Let us assume that $\Gamma$ is the subgroup of $G$ that is gauged on the Coulomb branch of the former quiver in order to construct the latter descending quiver. $\Gamma = G/H$ quotient corresponds to the amount of permutation symmetry that is lost by gauging a discrete subgroup of the global permutational symmetry of the parent quiver. Graph theoretically, it corresponds to the difference of permutation symmetry between the parent $\mathcal{P}_{[\lambda]}(n)$ bouquet and the descendant $\mathcal{P}_{[\chi]}(n)$ bouquet.  The ratio of the Coulomb branch volumes is shown in the upper-diagonal part of Table \eref{tab:RatiosGEN}. The ratios below the diagonal are inverse values of those above. Higher order Abelian and non-Abelian discrete groups that naturally show up in orbifold actions on bouquets quivers, as well as products of two or more quotient groups of the form $H_1 \times H_2$, are not included in the Table \eref{tab:RatiosGEN} for brevity. For example, the Coulomb branch of a quiver with a bouquet of two adjoint rank $2$ nodes, denoted by $\mathcal{C}_{[2^2]}$, is a $\mathbb{Z}_2 \times \mathbb{Z}_2$ orbifold of the a parent Coulomb branch $\mathcal{C}_{[1^4]}$, corresponding to a quiver with a complete bouquet of four rank $1$ nodes. We encounter such case in section \ref{1}. The Coulomb branches satisfy Equation \eref{eq:ratio}.
\be\label{eq:ratio}
\frac{vol(\mathcal{C}_{[1^4]})}{vol(\mathcal{C}_{[2^2]})} = ord(\mathbb{Z}_2) \times ord(\mathbb{Z}_2) =  2 \times 2 =4 .
\ee
\begin{table}
\begin{center}
\begin{tabular}{ |p{1.6cm}||p{1.3cm}|p{1.3cm}|p{1.3cm}|p{1.3cm}|p{1.3cm}|p{1.3cm}|p{1.3cm}|}
 \hline
 \multicolumn{8}{|c|}{Ratios of Quotients} \\
 \hline
 $G/H$ &$\mathbb{I}$&$\mathbb{Z}_2$&$\mathbb{Z}_3$&$S_3$&$S_4$&$S_5$&$S_6$\\
 \hline \hline
 $\mathbb{I}$&1&2&3&6&24&120&720\\
 $\mathbb{Z}_2$&&1&3/2&3&12&60&360\\
$\mathbb{Z}_3$&&&1&2&8&40&260\\
 $S_3$&&&&1&4&20&120\\
 $S_4$&&&&&1&5&30\\
 $S_5$&&&&&&1&6 \\
 $S_6$&&&&&&&1 \\
 \hline
\end{tabular}
\end{center}
\caption{\label{tab:RatiosGEN} Ratios of $t=1$ Hilbert Series poles for $k=3$, $n=3$ family}
\end{table}

\newpage

\section{Conclusions and Discussion} \label{5}

By the formulation of Conjecture \eref{Main Conjecture} we introduce general construction for discrete gauging in Coulomb branches of $3d$ $\CN=4$ quiver gauge theories. Although the construction is purely local and hence applicable to any quiver with a bouquet, it is demonstrated for three particular families of simply-laced bouquet quivers. The first and the third family are of particular interest since they serve as indispensable Coulomb branch tools for understanding Higgs branch phases of the $6d$ $\mathcal{N}=(1,0)$ world-volume theories of a stack of $n$ M5 branes on the $\mathbb{C}^2 / \mathbb{Z}_k$ singularity in M-theory.
In section \ref{4} the discrete gauging construction is extended to include a non-simply-laced quiver with a $C_2$ factor in the global symmetry. Central part of the work in this paper, concerning the first family of quivers, aims to offer a detailed analysis in support of Conjecture \eref{Main Conjecture}. The remaining part of this paper concerns the unrefined analysis of the quivers such that the ratios of the Coulomb branch volumes, defined in Equation \eref{eq:01}, are used as a non-trivial verification of Equation \eref{eq:00}.  As a remark, a complementary perspective on discrete gauging and its manifestation as discrete quotients on Coulomb branches is presented in \cite{HSdg18}.\\

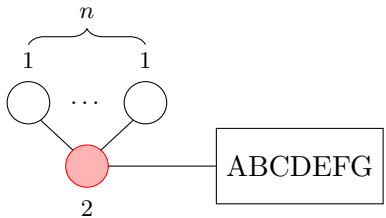
\begin{figure}[h!]
\center{
\begin{tikzpicture}[scale=0.70]
\draw (0,0)[red,fill=red!30] circle (0.4cm);
\draw (0,-0.8) node {\footnotesize{$2$}};
\draw (0.4,0) -- (2.425,0);
\draw (-0.26,0.3) -- (-0.87,0.87);
\draw (-1.1,1.2) circle (0.4cm);
\draw (-1.1,2) node {\footnotesize{$1$}};
\draw (0.26,0.3) -- (0.87,0.87);
\draw (1.1,1.2) circle (0.4cm);
\draw (1.1,2) node {\footnotesize{$1$}};
\draw (0,1.2) node {\footnotesize{$\dots$}};
\draw [decorate,decoration={brace,amplitude=6pt}] (-1.1,2.3) to (1.1,2.3);
\draw (0,2.9) node {\footnotesize{$n$}};
\node at (4,0) [minimum size=1cm,rectangle,draw] {ABCDEFG};
\end{tikzpicture}
}
\caption{\label{fig:ABCDEFGGen} $\mathcal{P}_{[1^{n}]}(n)$ Quiver with $SU(2)^{n}\times G$ global symmetry, where $G$ is any Lie group.}
\end{figure}

Possibly, an analogue of the general Formula \eref{eq:Gen} exists for other types of bouquet quivers. Consider a quiver that consists of two parts:
\begin{itemize}
\item A bouquet that stems from a rank $2$ unbalanced node
\item A second part, connected to the rank $2$ unbalanced node, that is itself a balanced ABCEFG Dynkin diagram.
\end{itemize}
Such quivers can be constructed by attaching a bouquet via the rank $2$ unbalanced node to a minimally unbalanced quiver. Quivers constructed in this manner take the form schematically depicted in Figure \eref{fig:ABCDEFGGen}.
The box on the right in Figure \eref{fig:ABCDEFGGen} symbolizes the balanced part of a minimally unbalanced quiver. The classification of all minimally unbalanced quivers is developed in \cite{MU18}. Recalling the general Formula \eref{eq:Gen} in section \ref{3}, one can speculate that the HWG for quivers of the form in Figure \eref{fig:ABCDEFGGen} involves:
\begin{itemize}
\item Order $t^2$: Adjoint representations of the balanced sub-quivers (i.e. $n$ copies of $SU(2)$ adjoint rep and a single adjoint rep corresponding to the balanced ABCDEFG part of the quiver). Let $\nu_i, i=1, ...,n$ denote the fugacities of the highest weights of the $n$ $SU(2)$ representations.
\item Order $t^n + t^{n+2}$: $n$-fundamental representation of $SU(2)$ combined with the representation that corresponds to the node of the ABCDEFG Dynkin diagram that is connected to the red node (i.e. the vector node in the case of $D_n$). Lets denote the highest weight fugacity for the representation of this ABCDEFG Dynkin node by the $\mu_{unbal}$.
\item Order $t^4$:  $\mu_{unbal}^2$ contribution and the typical singlet contribution
\item Order $t^{10}$: Relation transforming under $(\nu_1 ... \nu_n\; \mu_{unbal})^2$
\end{itemize}
 Interesting feature of this conjecture lies in the possibility to take moduli space with any particular isometry on the Coulomb branch and use discrete gauging to obtain various non-Abelian orbifolds of the original space. This is a novel method for constructing non-Aelian orbifold geometrical spaces with certain isometry. On the level of direct computation, however, it is challenging to obtain the HWG and the explicit verification of this conjecture is left for future study\footnote{The contribution appearing with the singlet at order $t^4$ is the most speculative part of the conjecture and requires verification.}. \\

The investigation of the analogue of Conjecture \eref{Main Conjecture} for ortho-symplectic quivers with bouquet nodes of type $O/Sp$ is one possible future direction. Another possible direction for development is the study of the same phenomenon in the context that involves M5 branes on an different type of singularity (i.e. the $D$-type or $E$-type singularities). Such analysis, however, is much more subtle due to the lack of intuition and complexity of the corresponding higher-dimensional physics.

 
\section*{Acknowledgments}
A.Z. would like to express a special gratitude to Rudolph Kalveks for indispensable help with some of the computations. A.Z. would also like to thank Santiago Cabrera and Marcus Sperling for useful and enlightening discussions. We would like to thank Julius Grimminger for pointing out a typo in \eref{eq:typo1}. A. H. would like to thank Ronen Plesser, Travis Maxfield, Gabi Zafrir, Santiago Cabrera, Rudolph Kalveks and Marcus Sperling for enlightening discussions. A.H. is supported by STFC Consolidated Grant ST/J0003533/1, and EPSRC Programme Grant EP/K034456/1.
\newpage

\appendix
\section{Construction of Bouquet Quivers} \label{A}

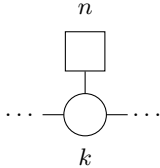
\begin{figure}[h!]
\center{
\begin{tikzpicture}[scale=0.70]
\draw (0.4,0) -- (0.8,0);
\draw (0,0) circle (0.4cm);
\draw (0,-0.8) node {\footnotesize{$k$}};
\draw (1.2,0) node {\footnotesize{$\dots$}};
\draw (-0.4,0) -- (-0.8,0);
\draw (-1.2,0) node {\footnotesize{$\dots$}};
\draw (0,0.4) -- (0,0.82);
\draw (0,2) node {\footnotesize{$n$}};
\node at (0,1.2) [minimum size=0.52cm,rectangle,draw] {};
\end{tikzpicture}
}
\caption{\label{fig:A1} Local part of a quiver with a $n$ flavor node attached to a $k$ gauge node.}
\end{figure}
\begin{figure}[h!]
\center{
\begin{tikzpicture}[scale=0.70]
\draw (0.4,0) -- (0.8,0);
\draw (0,0) circle (0.4cm);
\draw (0,-0.8) node {\footnotesize{$k$}};
\draw (1.2,0) node {\footnotesize{$\dots$}};
\draw (-0.4,0) -- (-0.8,0);
\draw (-1.2,0) node {\footnotesize{$\dots$}};
\draw (-0.26,0.3) -- (-0.87,0.87);
\draw (-1.1,1.2) circle (0.4cm);
\draw (-1.1,2) node {\footnotesize{$1$}};
\draw (0.26,0.3) -- (0.87,0.87);
\draw (1.1,1.2) circle (0.4cm);
\draw (1.1,2) node {\footnotesize{$1$}};
\draw (0,1.2) node {\footnotesize{$\dots$}};
\draw [decorate,decoration={brace,amplitude=6pt}] (-1.1,2.3) to (1.1,2.3);
\draw (0,2.9) node {\footnotesize{$n$}};
\end{tikzpicture}
}
\caption{\label{fig:A2} Local part of $\mathcal{P}_{[1^n]}(n)$ complete bouquet quiver.}
\end{figure}
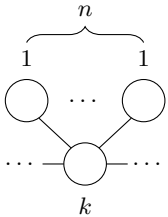

From a generic unitary $3d$ $\mathcal{N}=4$ quiver with a $n$ flavor node attached to a $k$ gauge node, one can obtain a \emph{complete bouquet} quiver by simply gauging the whole global symmetry into $n$ separate rank $1$ nodes. As an example, consider the quiver in Figure \eref{fig:A1}. Round and square nodes denote gauge and flavor groups, respectively.
To obtain the complete bouquet quiver, gauge the flavor node into separate $U(1)$ gauge nodes. The resulting quiver is shown in Figure \eref{fig:A2}. The form of the bouquet arrangement is denoted by $\mathcal{P}_{[1^n]}(n)$. This notation accordingly signifies that there are $n$ copies of rank $1$ nodes.  

The Coulomb brach of the quiver in Figure \eref{fig:A1}, denoted by $\mathcal{C}_1$, and the Coulomb brach of the quiver in Figure \eref{fig:A2}, denoted by $\mathcal{C}_2$ satisfy 
\be
\mathcal{C}_1 = \bigslant{\mathcal{C}_2 }{ U(1)^n}
\ee
where $\bigslant{}{}$ is used to denote a hyperK\"ahler quotient.

\section{Derivation of the HWG for the $\mathcal{P}_{[2^2]}$ theory} \label{B}

\begin{figure}[h!]
\center{
\begin{tikzpicture}[scale=0.70]
\draw (0,0) circle (0.4cm);
\draw (0,-0.8) node {\footnotesize{$Sp(1)$}};
\draw (0,0.4) -- (0,0.82);
\draw (0,2) node {\footnotesize{$D_4$}};
\node at (0,1.2) [minimum size=0.52cm,rectangle,draw] {};
\end{tikzpicture}
}
\caption{\label{fig:ab1} $Sp(1)$ gauge theory with $D_4$ global symmetry.}
\end{figure}
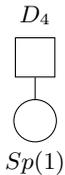

In this appendix we derive the HWG in Equation \eref{eq:hwg-22}. One starts with the HWG for the $Sp(1)$ gauge theory with $D_4$ flavor group, depicted in Figure \eref{fig:ab1} which has the form previously given in \eref{eq:hwg1}:
\be
HWG=PE[\mu_2 t^2].
\ee

Recall that under the action of the first $\mathbb{Z}_2$ the representation decomposition is given by \eref{eq:decomp1}. Hence, after the first $\mathbb{Z}_2$ action the HWG becomes
\be \label{eq:b2}
HWG=PE[\mu_2 t^2 +\mu_1^2 t^4],
\ee
where $\mu_i$ are the fugacities for the highest weights of $SO(7)$. In order to rewrite this HWG in terms of the $SU(4)$ fugacities, remember that the representations decompose as:
\begin{gather}
 \mu_2 \rightarrow \mu_1 \mu_3 + \mu_2 \\
\mu_1^2 \rightarrow \mu_2^2  + \mu_2+1,
\end{gather}
where on the LHS the $\mu_i$ are the highest weight fugacities of $SO(7)$ and on the RHS the $\mu_i$ are the highest weight fugacities of $SU(4)$, respectively. Thus, in terms of $SU(4)$, the HWG \eref{eq:b2} can be written in the form:
\be
HWG=PE[\mu_1 \mu_3 t^2+\mu_2t^2  +\mu_2^2t^4+\mu_2 t^4 +t^4 -\mu_2^2 t^8]
\ee
Note, that $\mu_2^2 t^8$ is subtracted to account for the undesired product of $\mu_2^2t^4$ and $t^4$. The $\mu_2$ transforms under the $\mathbb{Z}_2$ action with a minus sign therefore it must come in a form of a natural invariant $\mu_2^2 \;t^2$. Therefore, under the $\mathbb{Z}_2$ action, the HWG takes the form:
\be
HWG=PE[\mu_1 \mu_3 t^2+\mu_2^2 t^4  +\mu_2^2t^4+ \mu_2^2 t^6 +t^4 + \mu_2^2 t^8  -\mu_2^2 t^8 - \mu_2^4 t^{12}]
\ee
In summary, after the $\mathbb{Z}_2 \times \mathbb{Z}_2$ action the final HWG takes the form:
\be
HWG=PE[\mu_1 \mu_3 t^2+({2\mu_2}^2  +1)t^4 + {\mu_2}^2 t^6 - {\mu_2}^4 t^{12}].
\ee

\bibliographystyle{JHEP}
\bibliography{ImplosionActions}

\end{document}